\documentclass[preprint]{iacrtrans}

\usepackage[utf8]{inputenc}

\usepackage[T1]{fontenc}
\usepackage[utf8]{inputenc}
\usepackage{inputenc}
\usepackage[english]{babel}

\usepackage{hyperref} 
\usepackage{multirow}

\usepackage[ruled,linesnumbered]{algorithm2e}
\usepackage{amsmath,amssymb,amsfonts,amsthm}
\usepackage{tikz}
\usepackage{csquotes}
\usepackage{dashbox}
\usepackage{todonotes}
\usepackage{enumitem}   

\usepackage{graphicx,subfigure}
\definecolor{bblue}{HTML}{4F81BD}
\definecolor{rred}{HTML}{C0504D}
\definecolor{ggreen}{HTML}{9BBB59}
\definecolor{ppurple}{HTML}{9F4C7C}
\usepackage{xcolor}
\usepackage{tikz}
\usepackage{pgfplots}

\definecolor{findOptimalPartition}{HTML}{D7191C}
\definecolor{storeClusterComponent}{HTML}{FDAE61}
\definecolor{dbscan}{HTML}{ABDDA4}
\definecolor{constructCluster}{HTML}{2B83BA}

\usepackage{listings}
\usepackage{trace}
\usepackage{color}
\usepackage{makeidx}
\usepackage{mdframed}

\usepackage{listings}
\usepackage{color}
\usepackage{footmisc} 
\newcommand{\comment}[1]{}

\usepackage{fancyhdr} 
\usepackage{graphicx}

\newcommand{\qw}[1]{\textcolor{blue}{QW: #1}}
\usepackage{indentfirst}
 
\newtheorem{thm}{Theorem}
\newtheorem{defi}{Definition}

\newtheorem{prf}{Proof}

\usepackage{booktabs}    

\usepackage{dashbox}
\usepackage{tikz}

\usepackage{ulem}

\usepackage[
	n,
	operators,
	advantage,
	sets,
	adversary,
	landau,
	probability,
	notions,	
	logic,
	ff,
	mm,
	primitives,
	events,
	complexity,
	asymptotics,
	keys]{cryptocode}

\usepackage[most]{tcolorbox}

\author{Rujia Li\inst{1,2}$^{\dagger}$ \and Qin Wang\inst{3,4}\thanks{These authors contributed equally to the work.} \and Xinrui Zhang\inst{5} \and \\ Qi Wang\inst{1}\thanks{Corresponding author.} \and David Galindo\inst{2} \and Yang Xiang\inst{3}}
\institute{\textit{Southern University of Science and Technology}, Shenzhen, 518055, Guangdong, China \and \textit{University of Birmingham}, Edgbaston, B15 2TT, Birmingham, United Kingdom. \email{} \and \textit{Swinburne University of Technology}, Melbourne, VIC 3122, Australia. \and \textit{CSIRO Data61}, Sydney, NSW 2015, Australia. \and \textit{Nankai University}, Tianjin, 300350, China. \\ \email{rxl635@bham.ac.uk, qinwang@swin.edu.au, wangqi@sustech.edu.cn} }

\title[\texttt{iacrtans} class documentation]{An Offline Delegatable Cryptocurrency System}
\subtitle{(Full Version)\thanks{A preliminary version was accepted by IEEE ICBC'21~\cite{rujia2021delega}. A poster version was shown in NDSS'21~\cite{poster2021}.}}

\begin{document}

\maketitle
% use optional argument because the \LaTeX command breaks the PDF keywords
\keywords[\publname, TCHES, LaTeX]{Cryptocurrency, Payment, Trusted Execution Environments (TEEs), Offline Delegation}

\begin{abstract}
 Blockchain-based cryptocurrencies, facilitating the convenience of payment by providing a decentralized online solution, have not been widely adopted so far due to slow confirmation of transactions. Offline delegation offers an efficient way to exchange coins. However, in such an approach, the coins that have been delegated confront the risk of being spent twice since the delegator's behaviour cannot be restricted easily on account of the absence of effective supervision. Even if a third party can be regarded as a judge between the delegator and delegatee to secure transactions, she still faces the threat of being compromised or providing misleading assure. Moreover, the approach equipped with a third party contradicts the real intention of decentralized cryptocurrency systems. In this paper, we propose \textit{DelegaCoin}, an offline delegatable cryptocurrency system to mitigate such an issue. We exploit trusted execution environments (TEEs) as decentralized ``virtual agents'' to prevent malicious delegation. In DelegaCoin, an owner can delegate his coins through offline-transactions without interacting with the blockchain network. A formal model and analysis, prototype implementation and further evaluation demonstrate that our scheme is provably secure and practically feasible.

\end{abstract}

\newpage

%=====================================
\section{Introduction}
%=====================================

The interest in decentralized cryptocurrencies has grown rapidly in recent years. Bitcoin \cite{nakamoto2008bitcoin}, as the first and most famous system, has attracted massive attention. Subsequently, a handful of cryptocurrencies, such as Ethereum \cite{wood2014ethereum}, Namecoin \cite{ali2016blockstack} and Litecoin \cite{reed2017litecoin}, were proposed. Blockchain-based cryptocurrencies significantly facilitate the convenience of payment by providing a decentralized online solution for customers. However, merely online processing of transactions confronts the problem of low performance and high congestion. Offline delegation provides an alternative way to mitigate the issue by enabling users to exchange the coin without having to connect to an online blockchain platform~\cite{gudgeon2020sok}. Unfortunately, decentralized offline delegation still confronts risks caused by unreliable participants. The misbehaviours may easily happen due to the absence of effective supervision. To be specific, let us start from a real scenario: imagine that Bob, the son of Alex, a wild teenager, wants some digital currency (\textit{e.g.}, BTC) to buy a film ticket. According to current decentralized cryptocurrency payment technologies \cite{nakamoto2008bitcoin}\cite{wood2014ethereum}, Alex has two delegation approaches: (1) \textit{Coin-transfer.} Alex asks for Bob's BTC address, and then transfers a specific amount of coins to Bob's address. In such a scenario, Bob can only spend received coins from Alex. (2) \textit{Ownership-transfer.} Alex directly gives his own private key to Bob. Then, Bob can freely spend the coins using such a private key. In this situation, Bob obtains all coins that are saved in Alex's address.

We observe that both approaches suffer drawbacks. For the first approach, coin-transfer requires a global consensus of the blockchain, which makes it time-consuming \cite{kiayias2015speed}. For example, a confirmed transaction in the Bitcoin \cite{nakamoto2008bitcoin} takes around one hour (6 blocks), making the coin-transfer lose the essential property of real-time. For the other approach, ownership-transfer highly relies on the honesty of the delegatee. The promise between the delegator and delegatee depends on their trust or relationship. But it is weak and unreliable. The delegatee may spend all coins in the address for other purposes. Back to the example, Alex's original intention is to give Bob 200 $\mu BTC$ to buy a film ticket, but Bob may spend all coins to purchase his favorite toys. That means Alex loses control of the rest of coins. These two types of approaches represent most of the mainstream schemes ever aiming to achieve a secure delegation, but neither of them provide a satisfactory solution. This leads to the following research problem:

\begin{center}
\begin{tcolorbox}[colback=gray!10,%gray background
                  colframe=black,% black frame colour
                  width=12cm,% Use 8cm total width,
                  arc=1mm, auto outer arc,
                  boxrule=0.5pt,
                 ]
Is it possible to build a secure offline peer-to-peer delegatable system for decentralized cryptocurrencies?
\end{tcolorbox}
\end{center}

\noindent The answer would intuitively be ``NO''. Without interacting with the online blockchain network, the coins that have been used confront the risk of being spent twice after another successful delegation. This is because a delegation is only witnessed by the owner and delegatee, where no authoritative third parties perform final confirmation. The pending status leaves a window for attacks in which a malicious coin owner could spend this delegated transaction before the delegatee uses it. Even if a third party can be introduced as a judge between the delegator (owner) and delegatee to secure transactions, she faces the threat of being compromised or provided with misleading assure. Furthermore, the approach equipped with a third party contradicts the real intention of decentralized cryptocurrency systems.

In this paper, we propose \textit{DelegaCoin}, an offline delegatable electronic cash system. The trusted execution environments (TEEs) are utilized to play the role of \textit{virtual agent}. TEEs prevent malicious delegation of the coins (\textit{e.g.} double-delegation on the same coins). As shown in Figure.\ref{size}, the proposed scheme allows the owner to delegate her coins without interacting with the blockchain or any trusted third parties. The owner is able to directly delegate specific amounts of coins to others by sending them through a secure channel. This delegation can only be executed once under the supervision of delegation policy inside TEEs. In a nutshell, this paper makes the following contributions.

\begin{itemize}
    \item[-] We propose an offline delegatable payment solution, called \textit{DelegaCoin}. It employs the trusted execution environments (TEEs) as the decentralized \textit{virtual agents} to prevent the malicious owner from delegating the same coins multiple times. 
    
    \item[-] We formally define our protocols and provide a security analysis. Designing a provably secure system from TEEs is a non-trivial task that lays the foundation for many upper-layer applications. The formal analysis indicates that our system is secure.
    
    \item[-] We implement the system with Intel’s Software Guard Extensions (SGX) and conduct a series of experiments including the time cost for each function and the used disk space under different configurations. The evaluations demonstrate that our system is feasible and practical.
    
\end{itemize}

\smallskip
\noindent\textbf{Paper Structure.} Section~\ref{sec-rw} gives the background and related studies. Section~\ref{sec-prelimi} provides the preliminaries and building blocks. Section~\ref{sec-design} outlines the general construction of our scheme. Section~\ref{sec-formal} presents a formal model for our protocols.  Section~\ref{sec-seurity} provides the corresponding security analysis. Section~\ref{sec-implementation} and Section~\ref{sec-evaluation} show our implementation and evaluation, respectively. Section~\ref{sec-conclusion} concludes our work. Appendix A provides an overview of protocol workflow, Appendix B shows the resources availability and Appendix C presents featured notations in this paper.

%====================================================
\section{Related Work}
\label{sec-rw}
%====================================================

\noindent\textbf{Decentralized Cryptocurrency System.}
Blockchain-based cryptocurrencies facilitate the convenience of payment by providing a decentralized online solution for customers. Bitcoin \cite{nakamoto2008bitcoin} was the first and most popular decentralized cryptocurrency. Litecoin \cite{reed2017litecoin} modified the PoW by using the Script algorithm and shortened the block confirmation time. Namecoin \cite{ali2016blockstack} was the first hard fork of Bitcoin to record and transfer arbitrary names (keys) securely. Ethereum \cite{wood2014ethereum} extended Bitcoin by enabling state-transited transactions. Zcash \cite{hopwood2016zcash} provides a privacy-preserving payment solution by utilizing zero-knowledge proofs. CryptoNote-style schemes \cite{van2013cryptonote}, instead, enhance the privacy by adopting ring-signatures. However, slow confirmation of transactions retards their wide adoption from developers and users. Current cryptocurrencies, with ten to hundreds of TPS~\cite{nakamoto2008bitcoin,zheng2018detailed}, cannot rival established payment systems such as Visa or PayPal that process thousands. Thus, various methods have been proposed for better throughput. The scaling techniques can be categorized in two ways: (i) On-chain solutions that aim to create highly efficient blockchain protocols, either by reconstructing structures \cite{wang2020sok}, connecting chains \cite{zamyatin2019sok} or via sharding the blockchain \cite{wang2019sok}. However, on-chain solutions are typically not applicable to existing blockchain systems (require a hard fork). (b) Off-chain (layer 2) solutions that regard the blockchain merely as an underlying mechanism and process transactions offline  \cite{gudgeon2020sok}. Off-chain solutions generally operate independently on top of the consensus layer of blockchain systems, not changing their original designs. In this paper, we explore the second avenue.

\smallskip
\noindent\textbf{TEEs and Intel SGX.} 
The Trusted Execution Environments (TEEs) provide a secure environment for executing code to ensure the confidentiality and integrity of code and logic \cite{ekberg2013trusted}. State-of-the-art implementations include Intel Software Guard Extensions (SGX)~\cite{costan2016intel}, ARM TrustZone~\cite{pinto2019demystifying}, AMD memory encryption~\cite{kaplan2016amd}, Keystone ~\cite{lee2020keystone}, \textit{etc}. Besides, many other applications like BITE \cite{Matetic2018BITEBL}, Tesseract \cite{bentov2019tesseract}, Ekiden \cite{cheng2019ekiden} and Fialka \cite{li2020accountable} propose their TEEs-empowered schemes, but they still miss the focus of offline delegation. In this paper, we utilize SGX \cite{costan2016intel} to construct the system. SGX is one of TEEs representatives, and offers a set of instructions embedded in central processing units (CPUs). These instructions are used for building and maintaining CPUs' security areas. To be specific, SGX allows to create private regions (\textit{a.k.a.} enclave) of memory to protect the inside contents. The following features are highlighted in this technique: (1) \textit{Attestation.} Attestation mechanisms are used to prove to a validator that the enclave has been correctly instantiated, and used to establish a secure, authenticated connection to transfer sensitive data. The attestation guarantees the secret (private key) to be provisioned to the enclave only after a successful substantiation. (2) \textit{Runtime Isolation.} Processes inside the enclave are protectively isolated from the software running outside. Specifically, the enclave prevents higher privilege processes and outside operating system codes from falsifying inside executions of loaded codes. (3) \textit{Sealing identity technique.} SGX offers a seal sealing identity technique, where the enclave data is allowed to store in untrusted disk space. The private sealing key comes from the same platform key, which enables data sharing across different enclaves.

\smallskip
\noindent\textbf{Payment Delegation.} The payment delegation plays a crucial role in e-commercial activities, and it has been comprehensively studied for decades. Several widely adopted approaches are such that using credit cards (Visa, Mastercard, etc.), using reimbursement, using third-party platforms (like PayPal~\cite{williams2007introduction}, AliPay~\cite{guo2016ecosystem}). These schemes allow users to delegate their cash spending capability to their own devices or other users. However, these delegation mechanisms heavily rely on a centralized party that needs a fairly great amount of trust. Decentralized cryptocurrencies, like Bitcoin \cite{nakamoto2008bitcoin} and Ethereum \cite{wood2014ethereum}, remove the role of trusted third parties, making the payment reliable and guaranteed by distributed blockchain nodes. However, such payment is time-consuming since the online transactions need to get confirmed by the majority of participated nodes. The delegation provides the decentralized cryptocurrency with an efficient payment approach to delegate the coin owner's spending capability. The cryptocurrency delegation using SGX was first explored in \cite{matetic2018delegatee}, where they only focused on the credential delegation in the fair exchange. Teechan \cite{lind2019teechain} provided a full-duplex payment channel framework that employs TEEs, in which the parties can pay each other without interacting with the blockchain in a bounded time. However, Teechan requires a complex setup: the parties must commit a \textit{multisig} transaction before the channel started. In contrast, our scheme is simple and more practical.

%=====================================
\section{Preliminaries and Definitions} 
\label{sec-prelimi}
%=====================================

We make use of the following notions, definitions and assumptions to construct our scheme. Details are shown as follows.

\subsection{Notions}

Let $\mathsf{\lambda}$ denote a security parameter, $\mathsf{negl(\lambda)}$ represent a negligible function and $\mathcal{A}$ refer to an adversary. $\mathsf{b_{\star}}$ and $\mathsf{c_{\star}}$ are wildcard characters, representing the balance and the encrypted balance, respectively. A full notion list is provided in Appendix~\ref{appendix:b}.

%is the initial balance and $\mathsf{c_{\star}}$ is the initial encrypted balance. This step prevents unexpected occasions that may destroy the state in TEEs memory. 

%IND-CCA2 \cite{kiltz2003general} secure public key encryption scheme, $\mathsf{S}$ represent an existentially unforgeable (EUF-CMA) signature scheme~\cite{goldwasser1988digital}, and $\mathsf{SE}$ refer to a IND-CPA \cite{kaltz2008introduction} secure symmetric encryption scheme. Then, we provide our formal protocols in the following.

\subsection{Crypto Primitive Definitions}

\noindent \textbf{Semantically Secure Encryption.} A semantically secure encryption $\mathsf{SE}$ consists of a triple of algorithms $\mathsf{(KGen, Enc, Dec)}$ defined as follows.

%The key generation algorithm takes in a security parameter and outputs a key sk from the key space K

\begin{itemize}
\item[-] $\mathsf{SE.KGen}(1^\lambda)$ The algorithm takes as input security parameter $1^{\lambda}$ and generates a private key $\sk$ from the
key space $\mathcal{M}$.

\item[-] $\mathsf{SE.Enc(\sk, msg)}$ The algorithm takes as input a private key $\sk$ and a message $\mathsf{msg} \in \mathcal{M}$, and outputs a ciphertext $\mathsf{ct}$.

\item[-] $\mathsf{SE.Dec(\sk,ct)}$ The algorithm takes as input a verification key $\sk$, a message $\mathsf{ct}$, and outputs $\mathsf{msg}$.
\end{itemize}

\smallskip
\noindent\textit{Correctness}. A semantically secure encryption scheme $\mathsf{SE}$ is correct if for all $\mathsf{msg} \in \mathcal{M}$, 
\begin{align*}
\Pr\big[\mathsf{SE.Dec(\sk,(SE.Enc(\sk,msg))) \neq msg}  \big| \mathsf{\sk \gets SE.KGen(1^\lambda)}\big] \leq \mathsf{negl(\lambda)},
\end{align*}
where $ \mathsf{negl(\lambda)}$ is a negligible function and the probability is taken over the random coins of the algorithms $\mathsf{SE.Enc}$ and $\mathsf{SE.Dec}$.

\begin{defi}[IND-CPA security of $\mathsf{SE}$]\label{secpa}
A semantically secure encryption scheme $\mathsf{SE}$ achieves Indistinguishability under Chosen-Plaintext Attack(IND-CPA) if
all PPT adversaries, there exists a negligible function $\mathsf{negl(\lambda)}$ such that
\begin{align*}
\big| \Pr\big[ \mathsf{G_{\adv, SE}^{IND-CPA}(\lambda)} = 1\big] - \frac{1}{2} \big| \leq \mathsf{negl(\lambda)},
\end{align*}
where $\mathsf{G_{\adv, SE}^{IND-CPA}(\lambda)}$ is defined as follows: 

\begin{pcvstack}[center]%
   \procedure{$\mathsf{G_{\adv, SE}^{IND-CPA}(\lambda)}$}{%
    \pcln \mathsf{\sk} \stackrel{\$}{\leftarrow} \mathsf{SE.KGen}(1^\lambda);  \\
    \pcln \mathsf{b} \stackrel{\$}{\leftarrow} \{0,1\} \\
    \pcln \mathsf{m_{0},m_{1}} \gets \mathcal{A}^{\mathsf{SE}(\cdot)} \\
    \pcln \mathsf{c^\star} \gets \mathsf{SE.Enc(\sk,m_b}) \\
   \pcln \mathsf{b^{'}} \gets \adv^{\mathsf{SE}(\cdot)}\mathsf{(c^\star)}\\
   \pcln \pcreturn  \mathsf{b = b^{'}}  
   } 
\end{pcvstack}

\end{defi}

\noindent \textbf{Signature Scheme.} A signature scheme $\mathsf{S}$ consists of the following algorithms.

\begin{itemize}
\item[-] $\mathsf{S.KeyGen}(1^\lambda)$ The algorithm takes as input security parameter $1^{\lambda}$ and generates a private signing key $\sk$ and a public verification key $\vk$.

\item[-] $\mathsf{S.Sign(\sk, msg)}$ The algorithm takes as input a signing key $\sk$ and a message $\mathsf{msg} \in \mathcal{M}$, and outputs a signature $\mathsf{\sigma}$.

\item[-] $\mathsf{S.Verify(\vk,\sigma,msg)}$ The algorithm takes as input a verification key $\vk$, 
a signature $\mathsf{\sigma}$ and a message $\mathsf{msg} \in \mathcal{M}$, and outputs $1$ or $0$.
\end{itemize}

\smallskip
\noindent\textit{Correctness}. A signature scheme $\mathsf{S}$ is correct if for all $\mathsf{msg} \in \mathcal{M}$, 
\begin{align*}
\Pr\big[\mathsf{S.Verify(\vk,(S.Sign(\sk, msg)),msg)} \neq 1 \big| \mathsf{(\vk,\sk) \gets S.KeyGen(1^\lambda)}\big] \leq \mathsf{negl(\lambda)},
\end{align*}
where $ \mathsf{negl(\lambda)}$ is a negligible function and the probability is taken over the random coins of the algorithms $\mathsf{S.Sign}$ and $\mathsf{S.Verify}$.

\begin{defi}[(EUF-CMA security of $\mathsf{S}$]\label{secpa}
A signature scheme $\mathsf{S}$ is called Existentially Unforgeable under Chosen Message Attack(EUF-CMA) if all PPT adversaries, there exists a negligible function $\mathsf{negl(\lambda)}$ such that 
\begin{align*}
 \Pr\big[ \mathsf{G_{\adv, S}^{EUF-CMA}(\lambda)} = 1\big]  \leq \mathsf{negl(\lambda)},
\end{align*}
where $\mathsf{G_{\adv, S}^{EUF-CMA}(\lambda)}$ is defined as follows: 

\begin{pcvstack}[center]%
   \procedure{$\mathsf{G_{\adv, S}^{IND-CPA}(\lambda)}$}{%
    \pcln \mathsf{(\sk, pk)} \stackrel{\$}{\leftarrow} \mathsf{S.KeyGen}(1^\lambda);  \\
    \pcln \mathcal{L} \gets \mathsf{S.Sign(\sk, m_{\{0,\dots,n\}})};  \\
    \pcln \mathsf{(m^{\star},\sigma^{\star})} \gets \mathcal{A}^{\mathcal{O}(sk, \cdot)} \mathsf{(pk)}\\
   \pcln \pcreturn  \mathsf{( S.Verify(vk,\sigma^{\star}, m^{\star})} = 1) \wedge \mathsf{m^{\star} \notin \mathcal{L}}
   }
\end{pcvstack}

\end{defi}

\noindent \textbf{Public Key Encryption.} A public key encryption scheme $\mathsf{PKE}$ consists of the following algorithms.

\begin{itemize}
\item[-] $\mathsf{PKE.KeyGen}(1^\lambda)$ The algorithm takes as input security parameter $1^{\lambda}$ and generates a private signing key $\sk$ and a public verification key $\vk$.

\item[-] $\mathsf{PKE.Enc(pk, msg)}$ The algorithm takes as input in a public key $\mathsf{pk}$ and a message $\mathsf{msg} \in \mathcal{M}$, and outputs a ciphertext $\mathsf{ct}$.

\item[-] $\mathsf{PKE.Dec(\sk,ct)}$ The algorithm takes as input a secret key $\sk$, a ciphertext $\mathsf{ct}$, and outputs $\mathsf{msg}$ or $\bot$.

\end{itemize}

\smallskip
\noindent\textit{Correctness}. A public key encryption scheme $\mathsf{PKE}$ is correct if for all $\mathsf{msg} \in \mathcal{M}$, 
\begin{align*}
\Pr\big[\mathsf{SE.PKE(\sk,(PKE.Enc(pk,msg))) \neq msg}  \big| \mathsf{(\sk,pk) \gets PKE.KeyGen(1^\lambda)}\big] \leq \mathsf{negl(\lambda)},
\end{align*}
where $ \mathsf{negl(\lambda)}$ is a negligible function and the probability is taken over the random coins of the algorithms $\mathsf{PKE.KeyGen}$ and $\mathsf{PKE.Enc}$.

\begin{defi}[(IND-CCA2 security of  $\mathsf{PKE}$]\label{ccapke}
A PKE scheme $\mathsf{PKE}$ is said to have Indistinguishability Security under Adaptively Chosen Ciphertext Attack(IND-CCA2) if
all PPT adversaries, there exists a negligible function $\mathsf{negl(\lambda)}$ such that 
\begin{align*}
 \Pr\big[ \mathsf{G_{\adv,PKE}^{IND-CCA2}(\lambda)} = 1\big]  \leq \mathsf{negl(\lambda)},
\end{align*}
where $\mathsf{G_{\adv,PKE}^{IND-CCA2}(\lambda)}$ is defined as follows: 

\begin{pcvstack}[center]%
   \procedure{$\mathsf{G_{\adv, PKE}^{IND-CCA2}(\lambda)}$}{%
    \pcln \mathsf{(\sk, pk)} \stackrel{\$}{\leftarrow} \mathsf{PKE.KGen}(1^\lambda);  \\
    \pcln \mathsf{b} \stackrel{\$}{\leftarrow} \{0,1\} \\
    \pcln \mathsf{m_{0},m_{1}} \gets \mathcal{A}^{\mathsf{PKE.Dec}(\sk,\cdot)} \\
    \pcln \mathsf{c^\star} \gets \mathsf{PKE.Enc(\sk,m_b}) \\
   \pcln \mathsf{b^{'}} \gets \adv^{\mathsf{PKE.Dec}(\sk,\cdot)}\mathsf{(c^\star)}\\
   \pcln \pcreturn  \mathsf{b = b^{'}}  
   } 
\end{pcvstack}
\end{defi}

\subsection{Secure Hardware}
In our scheme, parties will have access to TEEs, in which they serve as isolated environments to guarantee the confidentiality and integrity of inside code and data. To capture the secure functionality of TEEs, inspired by~\cite{fisch2017iron,barbosa2016foundations} we define TEEs as a black-box program that provides some interfaces exposed to users. The abstraction is given as follows. Note that, Due to the scope of usage, we only capture the remote attestation of TEEs and refer to~\cite{fisch2017iron} for a full definition.

\begin{defi}

\label{TEEmode}
A secure hardware functionality $\mathsf{HW}$ for a class of probabilistic polynomial time (PPT) programs $\mathcal{P}$ includes the algorithms: $\mathsf{Setup}$, $\mathsf{Load}$, $\mathsf{Run}$, $\mathsf{Run Quote}$, $\mathsf{QuoteVerify}$. 

\begin{itemize}
\item[-] $\mathsf{HW.Setup(1^\lambda)}:$ The algorithm takes as input a security parameter $\lambda$, and outputs the secret key $\mathsf{sk_{quote}}$ and public parameters $\mathsf{pms}$.

\item[-] $\mathsf{HW.Load(pms}, P):$ The algorithm loads a stateful program $P$ into an enclave. It takes as input a program $P \in \mathcal{P}$ and $\mathsf{pms}$, and outputs a new enclave handle $\mathsf{hdl}$.

%used for the identification.

\item[-] $\mathsf{HW.Run(hdl,in)}:$ The algorithm runs enclave. It inputs a handle $\mathsf{hdl}$ that relates to an enclave (running program $P$) and an input $\mathsf{in}$, and outputs execution results $\mathsf{out}$.

\item[-] $\mathsf{HW.RunQuote(hdl, in)}:$ The algorithm executes programs in an enclave and generates an attestation quote. It takes as input $\mathsf{hdl}$ and  $\mathsf{in}$, and executes $P$ on $\mathsf{in}$. Then, it outputs $\mathsf{quote = (hdl,tag_P, in, out, \sigma)}$, where $\mathsf{tag_P}$ is a measurement to identify the program running inside an enclave and $\sigma$ is a corresponding signature.

\item[-] $\mathsf{HW.QuoteVerify(pms,quote)}:$ The algorithm verifies the quote. It firstly executes $P$ on $\mathsf{in}$ to get $\mathsf{out}$. Then, it takes as input $\mathsf{pms}$, $\mathsf{quote = (hdl,tag_P,in,out,\sigma)}$, and outputs $\mathsf{1}$ if the signature $\sigma$ is correct. Otherwise, it outputs $\mathsf{0}$.
\end{itemize}
\end{defi}

\smallskip
\noindent\textit{Correctness}. The $\mathsf{HW}$ scheme is correct if the following properties hold: For all program $\mathcal{P}$, all input $\mathsf{in}$

\begin{itemize}
\item Correctness of $\mathsf{HW.Run}$: for any specific  program $P \in \mathcal{P}$, the output of $\mathsf{HW.Run(hdl,in)}$ is deterministic.

\item Correctness of $\mathsf{RunQuote}$ and $\mathsf{QuoteVerify}$:
\begin{align*}
\Pr[\mathsf{QuoteVerify(pms, RunQuote} 
(\mathsf{hdl}, \mathsf{in})) \neq 1] \leq \mathsf{negl(\lambda).}\\
\end{align*}
\end{itemize}

Remote attestation in TEEs provides functionality for verifying the execution and corresponding output of a certain code run inside the enclave by using a signature-based quote. Thus, the remote attestation unforgeability security~\cite{fisch2017iron} is defined similarly to the unforgeability of a signature scheme.

\begin{defi}[Remote Attestation Unforgeability (RemAttUnf)]
\label{remoteAttestation} A $\mathsf{HW}$ scheme is RemAttUnf secure if all PPT adversaries, there exists a negligible function $\mathsf{negl(\lambda)}$ such that 
\begin{align*}
 \Pr\big[ \mathsf{G_{\adv, S}^{RemAttUnf}(\lambda)} = 1\big]  \leq \mathsf{negl(\lambda)},
\end{align*}
where $\mathsf{G_{\adv, HW}^{RemAttUnf}(\lambda)}$ is defined as follows:

\begin{pcvstack}[center]%
   \procedure{$\mathsf{G_{\adv, HW}^{RemAttUnf}(\lambda)}$}{%
    \pcln \mathsf{pms} \gets \mathsf{HW.Setup}(1^\lambda);  \\
    \pcln \mathsf{hdl} \gets \mathsf{HW.Load} (\mathsf{pms},P);  \\
    \pcln \mathcal{Q} \gets \mathsf{HW.RunQuote (hdl, in_{\{0,\dots,n\}})};  \\
    \pcln \mathsf{(in^{\star}, quote^{\star})} \gets \mathcal{A}^{\mathcal{O}(hdl,\cdot)} \mathsf{(pms)}\\
   \pcln \pcreturn  \mathsf{( HW.QuoteVerify(pms, quote^{\star})} = 1) \wedge \mathsf{quote^{\star} \notin \mathcal{Q}}
   } 
\end{pcvstack}

\end{defi}

%=====================================
\section{DelegaCoin}
\label{sec-design}
%=====================================

\begin{figure}[htb!]
\centering
\includegraphics[width=0.65\textwidth]{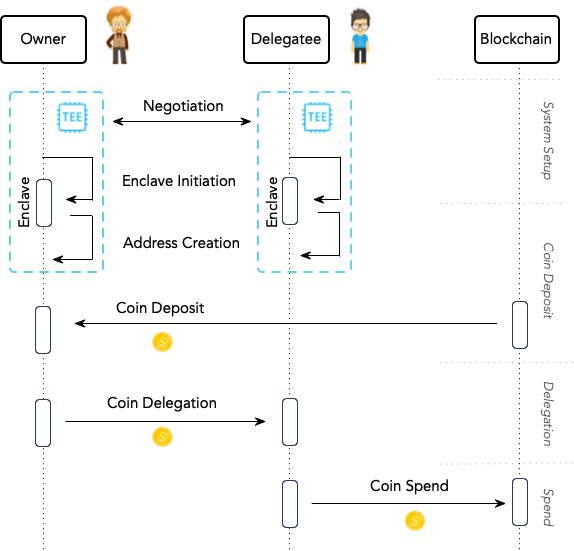}
\caption{DelegaCoin Workflow}
\label{size}
\end{figure}

In DelegaCoin, three types of entities are involved: coin owner (or delegator) $\mathcal{O}$, coin delegatee $\mathcal{D}$, and blockchain $\mathcal{B}$ (see Figure~\ref{size}). The main idea behind DelegaCoin is to exploit the TEEs as trusted agents between the coin owner and coin delegatee. TEEs are used to maintain delegation policies and ensure faithful executions of the delegation protocol. In particular, TEEs guarantee that the coin owner (either honest or malicious) cannot arbitrarily spend the delegated coins. The workflow is described as follows. Firstly, both $\mathcal{O}$ and $\mathcal{D}$ initialize and run the enclaves, and the owner $\mathcal{O}'s$ enclave generates an address $\mathsf{addr}$ for further transactions with a private key maintained internally. Next, $\mathcal{O}$ deploys delegation policies into the owner $\mathcal{O}'s$ enclave and deposits the coins to the address $\mathsf{addr}$. Then, $\mathcal{O}$ delegates the coins to $\mathcal{D}$ by triggering the execution of delegation inside the enclave. Finally, $\mathcal{D}$ spends delegated transaction to 
the blockchain network $\mathcal{B}$. Note that the enclaves in our scheme are decentralized, meaning that each  $\mathcal{O}$ and $\mathcal{D}$ has its own enclave without depending on a centralized agent, which satisfies the requirements of current cryptocurrency systems.

%----------------------------------------------
\subsection{System Framework}
%----------------------------------------------
    
\smallskip
\noindent\textbf{System Setup.}  In this phase, the coin owner $\mathcal{O}$ and the delegatee $\mathcal{D}$ initialize their TEEs to provide environments for the operations with respect to the further delegation.

\begin{itemize}
\item \textit{Negotiation.} $\mathsf{pms} \gets \mathsf{ParamGen(1^\lambda)}$: $\mathcal{O}$ agrees with $\mathcal{D}$ for the pre-shared information. Here, $\mathsf{\lambda}$ is a security parameter. 

\item \textit{Enclave Initiation.} $\mathsf{hdl}_\mathcal{O},\mathsf{hdl}_\mathcal{D} \gets \mathsf{EncvInit(1^\lambda,pms)}$: $\mathcal{O}$ and $\mathcal{D}$ initialize the enclave \textit{E}$_{\mathcal{O}}$ and \textit{E}$_{\mathcal{D}}$ with outputting the enclave handles $\mathsf{hdl}_\mathcal{O}$ and $\mathsf{hdl}_\mathcal{D}$.

\item \textit{Key Generation.}  $\mathsf{(pk_{Tx},sk_{Tx}),(pk_{\mathcal{O}},sk_{\mathcal{O}}), key_{seal}} \gets \mathsf{KeyGen^{TEE}}(\mathsf{hdl_{\mathcal{O}}},1^\lambda)$ and
$\mathsf{(pk_{\mathcal{D}}},\\ \mathsf{sk_{\mathcal{D}}}),(\mathsf{vk_{sign}}, \mathsf{sk_{sign}}), \mathsf{r} \gets \mathsf{KeyGen^{TEE}}(\mathsf{hdl_{\mathcal{D}}},1^\lambda)$: $\mathcal{O}$ and $\mathcal{D}$ run the enclaves \textit{E}$_{\mathcal{O}}$ and \textit{E}$_{\mathcal{D}}$ to create their internal keys. Key pair $\mathsf{(pk_{Tx},sk_{Tx})}$ is used for transaction generation. Key pair $(\mathsf{pk_{\mathcal{O}},sk_{\mathcal{O}}})$ and $(\mathsf{pk_{\mathcal{D}},sk_{\mathcal{D}}})$ are used for remote assertion, while $\mathsf{key_{seal}}$ is a sealing key used to export the state to the trusted storage. Key pair $(\mathsf{vk_{sign}}, \mathsf{sk_{sign}})$ is used to identify a specific delegatee, while $\mathsf{r}$ is a private key for transaction encryption.

\item \textit{Quote Generation.}  $\mathsf{ quote \gets QuoGen^{TEE}(\mathsf{sk_{\mathcal{O}}}, \mathsf{vk_{sign}}, pms)}$: $\mathcal{O}$ generate a $\mathsf{quote}$ for requesting an  encrypted symmetric encryption key from $\mathcal{D}$.

\item \textit{Key Provision.}  $\mathsf{ ct_{r} \gets Provision^{TEE}(quote,\mathsf{sk_{sign}}, \mathsf{pk_{\mathcal{O}}}, pms)}$: $\mathcal{O}$ proves to $\mathcal{D}$ that $\textit{E}_{\mathcal{O}}$ has been instantiated with a $\mathsf{quote}$ to request an encrypted symmetric encryption key $\mathsf{ct_{r}}$. The symmetric encryption is used to encrypt the messages inside TEEs.

\item \textit{Key Extraction.}  $\mathsf{ r \gets Extract^{TEE}(\mathsf{sk_{\mathcal{O}}}, ct_{r})}$: $\mathcal{O}$ extracts a symmetric encryption key $\mathsf{r}$ from $\mathsf{ct_{r}}$ using $\mathsf{sk_{\mathcal{O}}}$.

\item \textit{State Retrieval.}  $\mathsf{b_{init} = Dec^{TEE}(key_{seal}, c_{init})}$: Encrypted states are read back by the enclave \textit{E}$_{\mathcal{O}}$ under $\mathsf{key_{seal}}$, where $\mathsf{b_{init}}$ is the initial balance and $\mathsf{c_{init}}$ is the initial encrypted balance. This step prevents unexpected occasions that may destroy the state in TEEs memory. 

\end{itemize}

\smallskip
\noindent\textbf{Coin Deposit.} The enclave \textit{E}$_{\mathcal{O}}$ generates an address and its corresponding private key $\mathsf{pk_{Tx}}$ for the deposit. Afterwards, $\mathcal{O}$ sends coins to this address in the form of fund deposits.

\begin{itemize}

\item \textit{Address Creation.} $\mathsf{addr} \gets \mathsf{AddrGen^{TEE}}(1^\lambda,\mathsf{pk_{Tx}})$: $\mathcal{O}$ calls \textit{E}$_{\mathcal{O}}$ to generate a transaction address $\mathsf{addr}$. The private key $\mathsf{sk_{Tx}}$ of $\mathsf{addr}$ is secretly stored inside TEEs and is generated by an internal pseudo-random number.

\item \textit{Coin Deposit.} $\mathsf{ b_{deposit} = Update^{B}(addr,b_{init})}$: $\mathcal{O}$ generates an arbitrary transaction and transfers some coins to $\mathsf{addr}$ as the fund deposits.

\end{itemize}

%----------------------
\smallskip
\noindent\textbf{Coin Delegation.} In this phase, neither $\mathcal{O}$ nor $\mathcal{D}$ interacts with blockchain. $\mathcal{O}$ can instantly complete the coin delegation through offline transactions.

\begin{itemize}

\item \textit{Balance Update.} $\mathsf{b_{update} \gets Update^{TEE}(b_{deposit},b_{Tx})}$:  \textit{E}$_{\mathcal{O}}$ checks current balance to ensure that it is enough for deduction. Then, \textit{E}$_{\mathcal{O}}$ updates the balance.

\item \textit{Signature Generation.} $\mathsf{\sigma_{Tx}} \gets \mathsf{TranSign^{TEE}(\mathsf{sk_{Tx}},\mathsf{addr},b_{Tx})}$: \textit{E}$_{\mathcal{O}}$ generates a valid signature $\mathsf{\sigma_{Tx}}$.

\item \textit{Transaction Generation.} $\mathsf{Tx} \gets \mathsf{TranGen^{TEE}(\mathsf{addr},b_{Tx},\mathsf{\sigma_{Tx}})}$:  \textit{E}$_{\mathcal{O}}$ generates a transaction $\mathsf{Tx}$ using $\mathsf{\sigma_{Tx}}$.

\item  \textit{Coin Delegation.} $\mathsf{ct_{tx}} \gets \mathsf{TranEnc^{TEE}}(\mathsf{r},\mathsf{Tx})$: $\mathcal{O}$ sends encrypted transaction $\mathsf{ct_{tx}}$ to $\mathcal{D}$.

\item \textit{State Seal.} $\mathsf{c_{update} \gets Enc^{TEE}(key_{seal},b_{update})}$: Once completing the delegation, the records $\mathsf{c_{update}}$ are permanently stored outside the enclave. If any abort or halt happens, a re-initiated enclave starts to reload the missing information.

\end{itemize}
  
All the algorithms in the step of \textbf{Coin Delegation} must be run as an atomic operation, meaning that either all algorithms finish or none of them finish. 
A hardware Root of Trust can guarantee this, and we refer to~\cite{costan2016intel} for more detail.

\smallskip
\noindent\textbf{Coin Spend.} $\mathsf{Tx} \gets \mathsf{TranDec^{TEE}(r,\mathsf{ct_{tx}})}$:  $\mathcal{D}$ decrypts $\mathsf{ct_{tx}}$ with $\mathsf{r}$, and then spends $\mathsf{Tx}$ by forwarding it to blockchain network.

\smallskip
\noindent\textit{Correctness}. The DelegaCoin scheme is correct if the following properties hold:
For all $\mathsf{Tx}$, $\mathsf{b_{deposit}}$, $\mathsf{b_{update}}$ and $\mathsf{b_{Tx}}$.

\begin{itemize}
\item Correctness of $\mathsf{Update}$: 
$$\Pr \left[\mathsf{b_{Tx} \neq (b_{deposit} - b_{update})}\right] \leq \mathsf{negl(\lambda)}.$$

\item Correctness of $\mathsf{Seal}$:
$$\Pr[\mathsf{Dec^{TEE}(key_{seal},Enc^{TEE}(key_{seal}, b_{init})) \neq b_{init}}] \leq \mathsf{negl(\lambda)}.$$

\item Correctness of $\mathsf{Delegation}$:
\begin{align*}
\Pr[\mathsf{TranDec^{TEE}(r, TranEnc^{TEE}} 
(\mathsf{r}, \mathsf{Tx})) \neq \mathsf{Tx}] \leq \mathsf{negl(\lambda).}\\
\end{align*}
\end{itemize}

%============================
\subsection{Oracles for Security Definitions} 
\label{oracles}
%============================
We now define oracles to simulate an honest owner and delegatee for further security definitions and proofs. Each oracle maintains a series of (initially empty) sets $\mathcal{R}_1$, $\mathcal{R}_2$ and $\mathcal{C}$ which will be used later. Here, we use $\mathsf{(instruction; parameter)}$ to denote both the instructions and inputs of oracles.

\smallskip
\noindent\textbf{Honest Owner Oracle $\mathsf{O}^{\mathsf{owner}}:$} This oracle gives the adversary access to honest owners. An adversary $\mathcal{A}$ can obtain newly deletgated transactions or sealed storage with his customized inputs. The oracle provides the following interfaces.
\begin{itemize}

\item[-] On input $( \mathsf{signature\; creation}; \mathsf{addr})$, the oracle checks whether a tuple $(\mathsf{addr},\mathsf{\sigma_{Tx}}) \in \mathcal{R}_1$ exists, where $\mathsf{addr}$ is an input of transactions. If successful, the oracle returns $\mathsf{\sigma_{Tx}}$ to $\mathcal{A}$; otherwise, it computes $\mathsf{\sigma_{Tx}} \gets \mathsf{TranSign^{TEE}(\mathsf{sk_{Tx}}, \mathsf{addr},b_{Tx})}$ and adds $(\mathsf{addr},\mathsf{\sigma_{Tx}})$ to $\mathcal{R}_1$, and then returns $\mathsf{\sigma_{Tx}}$ to $\mathcal{A}$.

\item[-]  On input $(\mathsf{quote\; generation} ;\mathsf{vk_{sign}})$, the oracle checks if a tuple $(\mathsf{vk_{sign}},\mathsf{quote}) \in \mathcal{R}_2$ exists. If successful, the oracle returns $\mathsf{quote}$ to $\mathcal{A}$. Otherwise, it computes $\mathsf{quote} \gets \mathsf{QuoGen^{TEE}(sk_{\mathcal{O}}, vk_{sign}, pms})$ and adds $(\mathsf{vk_{sign}},\mathsf{quote})$ to $\mathcal{R}_2$, and then returns $\mathsf{quote}$ to $\mathcal{A}$.

\end{itemize}

\noindent\textbf{Honest Delegatee Oracle $\mathsf{O}^{\mathsf{delegatee}}:$} This oracle gives the adversary access to honest delegatees. The oracle provides the following interfaces.
\begin{itemize}

\item[-] On input $(\mathsf{key \;provision} ;\mathsf{quote})$, the oracle checks whether a tuple $(\mathsf{quote},\mathsf{ct_{r}}) \in \mathcal{C}$ exists. If successful, the oracle returns $\mathsf{ct_{r}}$ to $\mathcal{A}$; otherwise, it computes \\ $\mathsf{ ct_{r}  \gets Provision^{TEE}(quote,sk_{sign}, \mathsf{pk_{\mathcal{O}}}, pms)}$, adds $(\mathsf{quote},\mathsf{ct_{r}})$ to $\mathcal{C}$, and then returns $(\mathsf{quote},\mathsf{ct_{r}})$ to $\mathcal{A}$. 

\end{itemize}

\begin{figure}[htb!]
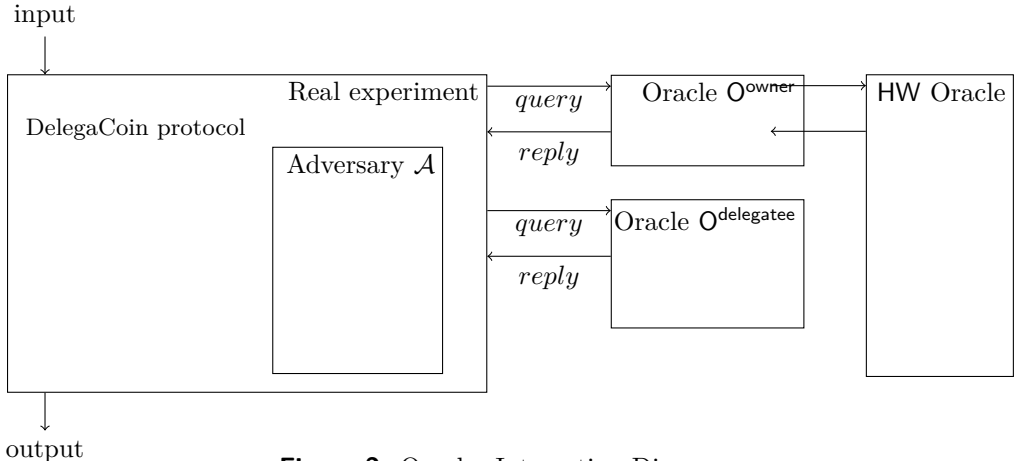

\centering
\caption{Oracles Interaction Diagram}
\begin{bbrenv}{A}
	\begin{bbrbox}[name=Real experiment]
	\pseudocode{
		\text{DelegaCoin protocol} 
	} %Reduction

	\begin{bbrenv}{B}
		\begin{bbrbox}[name=Adversary $\mathcal{A}$,minheight=3cm,xshift=3cm]
		\end{bbrbox}

	\end{bbrenv}

	\end{bbrbox}
	\bbrinput{input}
	\bbroutput{output}

	\begin{bbroracle}{OraA}
		\begin{bbrbox}[name=Oracle $\mathsf{O}^{\mathsf{owner}}$,minheight=1.2cm,minwidth=2.3cm]
		\end{bbrbox}
	\end{bbroracle}
	\bbroracleqryto{bottom=$query$}
	\bbroracleqryfrom{bottom=$reply$}

	\begin{bbroracle}{OraB}
		\begin{bbrbox}[name=Oracle $\mathsf{O}^{\mathsf{delegatee}}$,minheight=1.7cm,minwidth=2.3cm]
		\end{bbrbox}
	\end{bbroracle}
	\bbroracleqryto{bottom=$query$}
	\bbroracleqryfrom{bottom=$reply$}
	
	\begin{bbrbox}[name= $\mathsf{HW}$ Oracle ,minheight=4cm,xshift=11.3cm,minwidth=1.7cm]
	\end{bbrbox}
	\bbrmsgto{} %sidename=EndLoop
	\bbrmsgfrom{}
	%\bbroracleqryfrom{bottom=$m$}
	%\bbroracleqryto{bottom=$b$}
\end{bbrenv}
\end{figure}

\noindent\textbf{HW Oracle:} This oracle gives the adversary the access to honest hardware. 
The oracle provides the interfaces as defined as in~\ref{TEEmode}. Note that, to ensure that anything $\mathcal{A}$ sees in the real world can be simulated ideal experiment, we require that an adversary get access to \textbf{$\mathsf{HW}$ Oracle} through $\mathsf{O}^{\mathsf{delegatee}}$ and $\mathsf{O}^{\mathsf{owner}}$ rather than directly interact with $\mathsf{HW}$ Oracle.

%----------------------------------------------
\subsection{Threat Model and Assumptions}
%----------------------------------------------
As for involved entities, we assume that $\mathcal{O}$ attempts to delegate some coins to the delegatee. Each party may potentially be malicious. $\mathcal{O}$ may maliciously delegate an exceptional transaction, represented as sending the same transaction to multiple delegatees or spending the delegated transactions before the $\mathcal{D}$ spends them. $\mathcal{D}$ may also attempt to assemble an invalid transaction or double spend the delegated coins. We also assume the blockchain $\mathcal{B}$ is robust and publicly accessible.

With regard to devices, we assume that TEEs are secure, which means that an adversary cannot access the enclave runtime memory and their hardware-related keys (\textit{e.g.,} sealing key or attestation key). In contrast, we do not assume the components outside TEEs are trusted. For example, the adversary may control the operating system or high-level privileged software.

%------------------------------------

\subsection{Security Goals.}
%------------------------------------
DelegaCoin aims to employ TEEs to provide a secure delegatable cryptocurrency system. In brief, TEEs prevent malicious delegation in three aspects: (1) The private key of a delegated transaction and the delegated transaction itself are protected against the public. If an adversary learns any knowledge about the private key or the delegated transaction, she may spend the coin before the delegatee uses it; (2) The delegation executions are correctly executed. In particular, the spendable amount of delegated coins must be less than (or equal to) original coins; (3) The delegation records are securely stored to guarantee consistency considering accidental TEEs failures or malicious TEEs compromises. DelegaCoin is secure if adversaries cannot learn any knowledge about the private key, the delegated transaction, and the sealed storage.

%In the ideal world, the adversary sees a fake environment instead of the real one. 
To capture such security properties, we formalize our system through a game inspired by \cite{bernhard2015sok}. In our game, a PPT adversary attempts to distinguish between a real-world and a simulated (ideal) world. In the real world, the DelegaCoin algorithms work as defined in the construction. The adversary is allowed to access the transaction-related secret messages created by honest users through oracles as in Definition~\ref{oracles}. Obviously, the ideal world does not leak any useful information to the adversary. Since we model the additional information explicitly to respond to the adversary, we construct a polynomial-time simulator $\mathcal{S}$ that can \textit{fake} the additional information corresponding to the real result, but with respect to the fake TEEs. Thus, a universal oracle $\mathcal{U}(\cdot)$ in the ideal world is introduced to simulate the corresponding answers of $\adv$ called in oracles in the real world. We give a formal model as follows, in which these two experiments begin with the same setup assumptions.

\begin{defi}[Security]
DelegaCoin is simulation-secure if for all PPT adversaries $\mathcal{A}$, there exists a stateful PPT simulator $\mathcal{S}$ and a negligible function $ \mathsf{negl(\lambda)}$ such that the probability of that $\mathcal{A}$ distinguishes between $\mathsf{Exp_{\adv,{DelegaCoin}}^{real}(\lambda)}$ and $\mathsf{Exp_{\adv,{DelegaCoin}}^{idea}(\lambda)}$ is negligible, i.e.,
\begin{eqnarray}\nonumber
\left|\mathsf{Pr[Exp_{\adv,{DelegaCoin}}^{real}(\lambda)} = 1 ] - \mathsf{Pr[Exp_{\adv,{DelegaCoin}}^{ideal}(\lambda)} = 1 ] \right|  \leq \mathsf{negl(\lambda)}.
\end{eqnarray}
\end{defi}

\begin{figure}[htb]

\begin{pchstack}[center]
\resizebox{1.1\linewidth}{!}{
\fbox{
\begin{pcvstack}
\procedure{$\mathsf{Exp_{\adv,{DelegaCoin}}^{real}(\lambda)}$}{%
  \pcln \mathsf{pms} \gets \mathsf{ParamGen(1^\lambda)}  \\
   \pcln \mathsf{hdl_{\mathcal{O}}}, \mathsf{hdl_{\mathcal{D}}} \gets \mathsf{EncvInit(1^\lambda,pms)} \\
   \pcln \mathsf{(pk_{Tx}, sk_{Tx}), (\mathsf{pk_{\mathcal{O}}}, \mathsf{sk_{\mathcal{O}}}), key_{seal}} \gets \mathsf{KeyGen^{TEE}}(\mathsf{hdl_{\mathcal{O}}},1^\lambda) \\
   \pcln \mathsf{(pk_{\mathcal{D}},sk_{\mathcal{D}})},  (\mathsf{vk_{sign}}, \mathsf{sk_{sign}}), \mathsf{r} \gets \mathsf{KeyGen^{TEE}}(\mathsf{hdl_{\mathcal{D}}},1^\lambda) \\
   \pcln    \mathsf{ quote \gets \mathcal{A}(\mathsf{hdl_{\mathcal{O}}}, vk_{sign}, pms)} \\
   \pcln    \mathsf{ ct_{r} \gets \adv^{Provision^{TEE}(sk_{sign})}(\mathsf{hdl_{\mathcal{D}}}, quote, \mathsf{pk_{\mathcal{O}}}, pms)}  \\
   \pcln \mathsf{ r \gets \adv^{Extract^{TEE}(sk_{\mathcal{O}})}(\mathsf{hdl_{\mathcal{O}}}, ct_{r})} 
   \pclb 
   \pcintertext[dotted]{Setup Completed}
   \pcln \mathsf{b_{init} = Dec^{TEE}(\mathsf{hdl_{\mathcal{O}}},key_{seal}, c_{init})}\\
   \pcln \mathsf{addr} \gets \mathsf{AddrGen^{TEE}}(1^\lambda,\mathsf{pk_{Tx}})\\
   \pcln \mathsf{ b_{deposit} = Update^{B}(\mathsf{addr},b_{init})}\\
   \pcln \mathsf{b_{update} \gets Update^{TEE}(\mathsf{hdl_{\mathcal{O}}},b_{deposit},b_{Tx})} \\
   \pcln  \mathsf{\sigma_{Tx}} \gets \adv^{\mathsf{TranSign^{TEE}}(\mathsf{sk_{Tx}})}(\mathsf{hdl_{\mathcal{O}}}, \mathsf{addr,b_{Tx}})      \\
   \pcln \mathsf{Tx} \gets \mathsf{TranGen^{TEE}}(\mathsf{hdl_{\mathcal{O}}},\mathsf{addr,b_{Tx},\mathsf{\sigma_{Tx}}} )\\
   \pcln \mathsf{ct_{tx}} \gets \adv^{\mathsf{TranEnc^{TEE}(r)}}(\mathsf{hdl_{\mathcal{O}}}, \mathsf{Tx}) \\
   \pcln \mathsf{c_{update} = \adv^{Enc^{TEE}(key_{seal})}(\mathsf{hdl_{\mathcal{O}}},b_{update})} 
   \pclb
   \pcintertext[dotted]{Delegation Completed}
   \pcln \mathsf{Tx} \gets \mathsf{TranDec^{TEE}}(\mathsf{hdl_{\mathcal{D}}},\mathsf{r},\mathsf{ct_{tx}}) \\
   \pcln  \pcreturn (\mathsf{Tx},\mathsf{c_{update}})}
\end{pcvstack}
\pchspace
\procedure{$\mathsf{Exp_{\adv,{DelegaCoin}}^{ideal}(\lambda)}$}{%,\mathsf{c_{init}}
   \pcln \mathsf{pms} \gets \mathsf{ParamGen(1^\lambda)}  \\
   \pcln \mathsf{hdl_{\mathcal{O}}^\star}, \mathsf{hdl_{\mathcal{D}}^\star} \gets \mathsf{\mathcal{S}(1^\lambda,pms)} \\ 
   \pcln  \mathsf{(pk_{Tx},sk_{Tx}), (\mathsf{pk_{\mathcal{O}}}, \mathsf{sk_{\mathcal{O}}}),  key_{seal}} \gets \mathsf{KeyGen^{TEE}}(\mathsf{hdl_{\mathcal{O}}^{\star}},1^\lambda) \\
   \pcln \mathsf{(pk_{\mathcal{D}},sk_{\mathcal{D}})},(\mathsf{vk_{sign}}, \mathsf{sk_{sign}}), \mathsf{r} \gets \mathsf{KeyGen^{TEE}}(\mathsf{hdl_{\mathcal{D}}^{\star}},1^\lambda) \\
   \pcln    \mathsf{quote \gets \mathcal{A}(\mathsf{hdl_{\mathcal{O}}^{\star}},  vk_{sign}, pms)} \\
    \pcln    \mathsf{ ct_{r} \gets \adv^{\mathcal{S}^{\mathcal{U}(\cdot)}}(\mathsf{hdl_{\mathcal{D}}^{\star}}, quote,  \mathsf{pk_{\mathcal{O}}}, pms)} \\
    \pcln    \mathsf{ r \gets \adv^{\mathcal{S}^{\mathcal{U}(\cdot)}}(\mathsf{hdl_{\mathcal{O}}^{\star}},  ct_{r})} 
   \pclb 
   \pcintertext[dotted]{Setup Completed}
   \pcln \mathsf{b_{init}} \gets \mathsf{\mathcal{S}(\mathsf{hdl_{\mathcal{O}}},key_{seal}, c_{init})}  \\
   \pcln \mathsf{addr} \gets \mathcal{S}(1^\lambda,\mathsf{pk_{Tx}})\\
   \pcln \mathsf{b_{deposit} = \mathcal{S}(\mathsf{addr},b_{init})}\\
    \pcln \mathsf{b_{update} \gets \mathcal{S}(\mathsf{hdl_{\mathcal{O}}^\star},b_{deposit},1^{|b_{Tx}|})} \\
   \pcln  \mathsf{\sigma_{Tx}} \gets \mathsf{\adv^{\mathcal{S}^{\mathcal{U}(\cdot)}}(\mathsf{hdl_{\mathcal{O}}^\star}, \mathsf{addr},b_{Tx})}      \\
   \pcln \mathsf{Tx} \gets \mathcal{S}(\mathsf{hdl_{\mathcal{O}}^\star},\mathsf{addr}, \mathsf{1^{|b_{Tx}}|, \mathsf{\sigma_{Tx}}}) \\
   \pcln \mathsf{ct_{tx}} \gets \mathsf{\adv^{\mathcal{S}^{\mathcal{U}(\cdot)}}(\mathsf{hdl_{\mathcal{O}}^\star}, 1^{|Tx|})} \\
   \pcln \mathsf{c_{update} = \adv^{\mathcal{S}^{\mathcal{U}(\cdot)}}(\mathsf{hdl_{\mathcal{O}}^\star}, 1^{|b_{update}|})} 
   \pclb
   \pcintertext[dotted]{Delegation Completed}
   \pcln \mathsf{Tx} \gets \mathcal{S}(\mathsf{hdl_{\mathcal{D}}^\star},\mathsf{r},\mathsf{ct_{tx}}) \\
   \pcln \pcreturn (\mathsf{Tx},\mathsf{c_{update}})
 }}
 }
\end{pchstack}
\end{figure}

%=====================================
\section{Formal Protocols}
\label{sec-formal}
%=====================================

In this section, we present a formal model of our electronic cash system by utilizing the syntax of the $\mathsf{HW}$ model. In particular, we model the interactions of Intel SGX enclaves as calling to the $\mathsf{HW}$ functionality defined in Definition~\ref{TEEmode}. The formal protocols are provide as follows.

\smallskip
The owner enclave program $\mathsf{P_{\mathcal{O}}}$ is defined as follows. The value $\mathsf{tag_{P}}$ is a measurement of the program $\mathsf{P_{\mathcal{O}}}$, and it is hardcoded in the static data of $\mathsf{P_{\mathcal{O}}}$. Let $\mathsf{state}_{\mathcal{O}}$ denote an internal state variable.

$\mathsf{P_{\mathcal{O}}}$:
\begin{itemize}
\item On input (``init setup'', $\mathsf{sid, vk_{sign}}$\footnote{We assume that the combination ($\mathsf{sid}$,$\mathsf{vk_{sign}}$), represented as the identity of a delegatee, has already been distributed before the system setup. }):

    \begin{itemize}
    	\item[-] Run $(\mathsf{pk_{\mathcal{O}}}, \mathsf{sk_{\mathcal{O}}}) \gets \mathsf{\mathsf{PKE}.KeyGen}(1^\lambda)$ and $\mathsf{key_{seal}}\footnote{Multiple enclaves from the same signing authority can derive the same key, since seal key is based on the enclave’s certificate-based identity.} \gets \mathsf{\mathsf{SE}.KeyGen}(1^\lambda)$.
    	
    	\item[-] Update $\mathsf{state}$ to $(\mathsf{sk_{\mathcal{O}}, vk_{sign}})$ and output $(\mathsf{pk_{\mathcal{O}}, sid, vk_{sign}})$.
    	
    \end{itemize}
    
     \item On input (``complete setup'', $\mathsf{sid, ct_{r}, \sigma_{r}})$:
     
        \begin{itemize}
    	\item[-] Look up the $\mathsf{state}_{\mathcal{O}}$ to obtain the entry $\mathsf{(sk_{\mathcal{O}}, sid, vk_{sign})}$. If no entry
exists for $\mathsf{sid}$, output $\bot$.

    	\item[-] Receive the $(\mathsf{sid, vk_{sign}})$ from $\mathcal{O}$ and check if $\mathsf{vk_{sign}}$ matches with the one in $\mathsf{state}_{\mathcal{O}}$. If not, output $\bot$.

        \item[-] Verify signature $\mathsf{b \gets \mathsf{\mathsf{S}.Verify}(vk_{sign}, \sigma_{r}, (sid, ct_{r} ))}$.
        If $\mathsf{b}$ = 0, output $\bot$.
        
        \item[-] Run $\mathsf{r \gets \mathsf{\mathsf{PKE}.dec}(sk_{\mathcal{O}}, ct_{r})}$.
        
        \item[-] Add the tuple $(\mathsf{r, sid, vk_{sign}})$ to $\mathsf{state}_{\mathcal{O}}$.
    \end{itemize}
    
    \item On input (``state retrieval'', $\mathsf{sid}$):
    \begin{itemize}
        
    	\item[-] Retrieve identity-balance pair ($\mathsf{sid, c_{init}}$) from the sealed storage.
    	
    	\item[-] Run $\mathsf{b_{init} = \mathsf{SE}.Dec(key_{seal},c_{init})}$ and update $\mathsf{state}_{\mathcal{O}}$ to $(\mathsf{sid, b_{init}})$
    \end{itemize}
    
    \item On input (``address generation'', $1^\lambda$):
    \begin{itemize}
    	\item[-] Run $(\mathsf{sk_{Tx}}, \mathsf{pk_{Tx}}) \gets \mathsf{\mathsf{S}.KeyGen}(1^\lambda)$ and $\mathsf{addr} \gets \mathsf{AddrGen^{TEE}}(1^\lambda,\mathsf{pk_{Tx}})$.
    	\item[-] Update $(\mathsf{sk_{Tx}, addr})$ to $\mathsf{state}_{\mathcal{O}}$  and output $(\mathsf{pk_{Tx}}, \mathsf{addr})$.
    \end{itemize}
    
    \item On input (``transaction generation'', $\mathsf{addr}$ ):
    \begin{itemize}
        \item[-] Retrieve the private key $\mathsf{sk_{Tx}}$.
    	\item[-] Run $\mathsf{\sigma_{Tx}} \gets \mathsf{\mathsf{S}.Sign(sk_{Tx},\mathsf{addr},  b_{Tx})}$ and output a signature $\mathsf{\sigma_{Tx}}$.
    	\item[-] Run $\mathsf{Tx} \gets \mathsf{TranGen(\mathsf{addr},  b_{Tx},\mathsf{\sigma_{Tx}}})$ and update $(\mathsf{sid, Tx})$ to $\mathsf{state}_{\mathcal{O}}$.
    \end{itemize}
    
    \item On input (``state update'', $\mathsf{addr}$):
    \begin{itemize}
       \item[-] Check $\mathsf{b_{deposit}}$ and $\mathsf{b_{Tx}}$. If $\mathsf{b_{deposit} < b_{Tx}}$, output $\bot$.
    	\item[-] Run $\mathsf{b_{update} \gets Update(b_{deposit},b_{Tx})}$.
    \end{itemize}
    
    \item On input (``start delegation'', $\mathsf{addr}$):
    \begin{itemize}
    	\item[-] Retrieve the provision private key $\mathsf{r}$ and $\mathsf{Tx}$ from $\mathsf{state}_{\mathcal{O}}$.
    	
    	\item[-] Run $\mathsf{\mathsf{ct_{tx}} \gets \mathsf{SE}.\mathsf{Enc(r,Tx)}}$.   
    \end{itemize}
    
    \item On input (``state seal'', $\mathsf{addr}$):
    \begin{itemize}
    	\item[-] Run $\mathsf{\mathsf{c_{update}} = \mathsf{SE}.Enc(key_{seal},\mathsf{b_{update}})}$ and update $\mathsf{state}_{\mathcal{O}}$ to $(\mathsf{addr, b_{update}})$.
    	\item[-] Store $\mathsf{addr}$ and $\mathsf{c_{update}}$ to sealed storage.    
    \end{itemize}
    
\end{itemize}

\smallskip
The delegatee enclave program $\mathsf{P_{\mathcal{D}}}$ is defined as follows. The value $\mathsf{tag_{\mathcal{D}}}$ is the measurement of the program $\mathsf{P_{\mathcal{D}}}$, and it is hardcoded in the static data of $\mathsf{P_{\mathcal{D}}}$. Let $\mathsf{state}_\mathcal{D}$ denote an internal state variable. Also, the security parameter $\lambda$ is hardcoded into the program.

$\mathsf{P_{\mathcal{D}}}$:
\begin{itemize}
\item On input (``init setup'', $1^\lambda$):
    \begin{itemize}
    
       \item[-] Generate a session ID, $\mathsf{sid \gets \{0,1\}^{\lambda}}$. 
        
    	\item[-] Run $(\mathsf{pk_{\mathcal{D}}}, \mathsf{sk_{\mathcal{D}}}) \gets \mathsf{\mathsf{PKE}.KeyGen}(1^\lambda)$, and $(\mathsf{vk_{sign}}, \mathsf{sk_{sign}}) \gets \mathsf{\mathsf{S}.KeyGen}(1^\lambda)$. %\rj{$\mathsf{sk_{sign}}?$}
    	
    	\item[-] Update $\mathsf{state}_\mathcal{D}$ to $(\mathsf{sk_{\mathcal{D}},sk_{sign}})$ and output $(\mathsf{sid, pk_{\mathcal{D}},vk_{sign}})$.
    \end{itemize}
        
        \item On input (``provision'', $\mathsf{quote}, \mathsf{pk_{\mathcal{O}}}, \mathsf{pms}$):
    \begin{itemize}
    	\item[-] Parse $\mathsf{quote =(hdl_{\mathcal{O}}, tag_P, in, out, \sigma)}$,  check that $\mathsf{tag_{P}== tag_{\mathcal{O}}}$. If not, output $\bot$.
    
    	\item[-] Parse $\mathsf{out = (sid, pk_{\mathcal{O}})}$, and run $\mathsf{b \gets HW.QuoteVerify(pms,quote)}$ on $\mathsf{quote}$. If $\mathsf{b} = 0$, output $\bot$.
    	\item[-] Select a random number $\mathsf{r}$ and compute the algorithm
    	$\mathsf{ct_{r} = \mathsf{PKE}.Enc(pk_{\mathcal{O}},r)}$ and $\mathsf{\sigma_{r} = \mathsf{S}.Sign(sk_{sign}, (sid, ct_{r}))}$ and output $\mathsf{(sid, ct_{r}, \sigma_{r})}$.
    \end{itemize}
    
    \item On input (``complete delegation'', $\mathsf{ct_{tx}}$):
    \begin{itemize}
    	\item[-] Retrieve $\mathsf{r}$ from $\mathsf{state}_{\mathcal{D}}$.
    	\item[-] Run $\mathsf{\mathsf{Tx} \gets \mathsf{SE}.\mathsf{Dec(r,ct_{tx})}}$.   
    \end{itemize}
\end{itemize}

\smallskip
\noindent\hangindent 2em $\mathbf{Setup}.$ The following steps are based on the completed initialization of the programs of the delegator $\mathsf{P_{\mathcal{O}}}$ and delegatee $\mathsf{P_{\mathcal{D}}}$. The delegatee $\mathcal{D}$ runs $\mathsf{hdl_{\mathcal{D}}} \gets \mathsf{HW.Load(pms,P_{\mathcal{D}})}$ and $\mathsf{(\mathsf{vk_{sign}}, pk_{\mathcal{D}}) \gets HW.Run(hdl_{\mathcal{D}}, (\text{``init setup''}, 1^\lambda))}$. Then, $\mathcal{D}$ sends $\mathsf{vk_{sign}}$ to the delegator $\mathcal{O}$. Next, $\mathcal{O}$ runs $\mathsf{hdl}_{\mathcal{O}} \gets \mathsf{HW.Load(pms,P_{\mathcal{O}})}$ to load the handle. Meanwhile, $\mathcal{O}$ 
calls $\mathsf{quote \gets HW.Run\&Quote(hdl_{\mathcal{O}}, (\text{``init setup''}, \mathsf{sid, vk_{sign}}))}$, and sends a $\mathsf{quote}$ to $\mathcal{D}$. After that, $\mathcal{D}$ calls $\mathsf{(sid, ct_{r}, \sigma_{r})} \gets \mathsf{HW.Run(hdl_{\mathcal{D}}, (\text{``provision''}, \mathsf{quote,pk_{\mathcal{O}}, pms}))}$, and sends $\mathsf{(sid, ct_{r}, \sigma_{r})}$ to $\mathcal{O}$. Last, $\mathcal{O}$ 
calls $\mathsf{HW.Run(hdl_{\mathcal{O}}, (\text{``complete setup''}, \mathsf{vk_{sign}}))}$. At the end of completing setup, $\mathcal{O}$'s enclave \textit{E}$_{\mathcal{O}}$ obtains the private key $\mathsf{r}$ used for transaction delegation.

\smallskip
\noindent\hangindent 2em $\mathbf{Deposit}$. $\mathcal{O}$
calls $\mathsf{c_{init} \gets HW.Run(hdl_{\mathcal{O}}, (\text{``state retrieval''}, sid))}$. If $\mathsf{c_{init}}$ does not exist or equals to $0$, $\mathcal{O}$ calls
$\mathsf{addr \gets HW.Run(hdl_{\mathcal{O}}, (\text{``address generation''},1^\lambda))}$ to create a new address $\mathsf{addr}$. Then, $\mathcal{O}$ transfers some coins to $\mathsf{addr}$ through a normal blockchain transaction.

\smallskip
\noindent\hangindent 2em $\mathbf{Delegation}$. $\mathcal{O}$ firstly parses $\mathsf{hdl_{\mathcal{O}}}$ and
calls \textit{E}$_{\mathcal{O}}$. Then, \textit{E}$_{\mathcal{O}}$ retrieves the $\mathsf{addr}$. Afterwards, it calls $\mathsf{b_{update} \gets HW.Run(hdl_{\mathcal{O}}, (\text{``state update''},addr))}$. If the update algorithm returns false or failure, \textit{E}$_{\mathcal{O}}$ aborts the following operations. Otherwise, it looks up the state to obtain $\mathsf{sk_{Tx}}$, runs $\mathsf{Tx \gets HW.Run(hdl_{\mathcal{O}}, (\text{``transaction generation''}, addr ))}$ and outputs a transaction $\mathsf{Tx}$. After that, the delegator's enclave \textit{E}$_{\mathcal{O}}$ retrieves $\mathsf{r}$ and runs $\mathsf{ct_{tx} \gets HW.Run(hdl_{\mathcal{O}}, (\text{``start delegation''},addr))}$. Finally, ${\mathcal{O}}$ sends $\mathsf{ct_{tx}}$ to $\mathcal{D}$.

\smallskip
\noindent\hangindent 2em $\mathbf{Spend}$. $\mathcal{D}$ parses $\mathsf{hdl_{\mathcal{D}}}$ and runs $\mathsf{Tx \gets HW.Run(hdl_{\mathcal{D}}, (\text{``complete delegation''},\mathsf{ct_{tx}}))}$. After that, $\mathcal{D}$ spends the received transaction $\mathsf{Tx}$ by forwarding it to the blockchain network. Then, a blockchain node firstly parses 
$\mathsf{Tx = (addr,pk_{Tx},metadata,\sigma_{Tx})}$ and runs $\mathsf{b} \gets \mathsf{\mathsf{S}.Verify^{B}(pk_{Tx},\sigma_{Tx})}$. If $\mathsf{b} = 0$, output $\bot$. Otherwise, the node broadcasts $\mathsf{Tx}$ to other blockchain nodes.

%====================================================
\section{Security Analysis}
\label{sec-seurity}
%====================================================

\begin{thm}[Security]\label{prf-consistency-tee}
Assume that $\mathsf{\mathsf{SE}}$ is IND-CPA secure, 
$\mathsf{PKE}$ is IND-CCA2 secure, $\mathsf{S}$ holds the EUF-CMA security, and the TEEs are secure as in Definition~\ref{TEEmode}, DelegaCoin scheme is simulation-secure.
\end{thm}

Inspired by~\cite{lindell2017simulate, fisch2017iron}, we use a simulation-based paradigm to conduct security analysis, and explain the crux of our security proof as follows. We firstly construct a simulator $\mathcal{S}$ which can simulate the challenge responses in the real world. It provides the adversary $\adv$ with a simulated delegated transaction, a simulated quote and sealed storage. The information that $\adv$ can obtain is merely the instruction code and oracle responses queried by $\adv$ in the real experiment. At a high level, the proof idea is simple: $\mathcal{S}$ encrypts zeros as the challenge message. In the ideal experiment, $\mathcal{S}$ intercepts $\adv$'s queries to user oracle and provides simulated responses. It uses its $\mathcal{U(\cdot)}$ oracle to simulate oracles in the real world and sends the response back to $\adv$ as the simulated oracle output. $\mathcal{U(\cdot)}$ and $\mathcal{S}$'s algorithms are described as follows.

\smallskip
\noindent\textbf{Pre-processing phase.} $\mathcal{S}$ simulates the pre-processing phase similar to in the real world. It firstly runs $\mathsf{ParamGen(1^\lambda)}$ and records system parameters $\mathsf{pms}$ that are generated during the process. Then, it calls $\mathsf{EncvInit(1^\lambda,pms)}$ to create the simulated enclave instances.
$\mathcal{S}$ also creates empty lists $\mathcal{R}_1^\star$, $\mathcal{R}_2^\star$, $\mathcal{C}^\star$, $\mathcal{K}^\star$ and $\mathcal{L}^\star$ to be used later.

\smallskip
\noindent\hangindent 2em \smallskip
\noindent$\mathbf{KeyGen^{\star}(1^\lambda)}$ When $\mathcal{A}$ makes a query to \noindent$\mathbf{KeyGen(1^\lambda)}$ oracle, $\mathcal{S}$ responds the same way as in the
real world except that now stores all the public keys queried in a list $\mathcal{K}^{\star}$. That is, $\mathcal{S}$ does the following algorithms.

\begin{itemize}
\item[-] Compute and output $(\mathsf{pk_{\mathcal{O}}}, \mathsf{sk_{\mathcal{O}}}),(\mathsf{pk_{Tx}}, \mathsf{sk_{Tx}}) \gets \mathsf{\mathsf{PKE}.KeyGen}(1^\lambda)$.
\item[-] Store the keys $(\mathsf{pk_{\mathcal{O}}}, \mathsf{sk_{\mathcal{O}}}),(\mathsf{pk_{Tx}}, \mathsf{sk_{Tx}})$ in the list $\mathcal{K}^{\star}$.
\end{itemize}

\smallskip
\noindent\hangindent 2em \smallskip
\noindent$\mathbf{Enc^{\star}(key^\star, 1^{|{msg}^\star|})}$\footnote{Here, $\mathbf{msg^{\star}}$ is a wildcard character, representing any messages.} When $\mathcal{A}$ provides the challenge message $\mathsf{{msg}^\star}$ for symmetric encryption, the following algorithm is used by $\mathcal{S}$ to simulate the challenge ciphertext.

\begin{itemize}
\item[-] Compute and output $\mathsf{ct^\star \gets \mathsf{SE}.Enc(key^\star, 1^{|{msg}^\star|)}}$.
\item[-] Store $\mathsf{ct}^\star$ in the list $\mathcal{L}^{\star}$.
\end{itemize}

\smallskip
\noindent\hangindent 2em $\mathbf{\mathsf{O}^{owner\star}(\mathsf{signature\; creation;addr)}}.$ When $\mathcal{A}$ takes a query to $\mathbf{\mathsf{O}^{owner}}$ oracle, $\mathcal{S}$ responds the same way as in the real world, except that $\mathcal{S}$ now stores all the $\mathsf{addr}$ corresponding to the user's queries in a list $\mathcal{R}_1^\star$. That is, $\mathcal{S}$ does the following algorithms.

\begin{itemize}
\item[-]  Call $\mathbf{\mathsf{O}^{owner}}$ oracle with an input $( \mathsf{signature\; creation}; \mathsf{addr})$ and output $\mathsf{\sigma_{Tx}}$.
\item[-] Store $(\mathsf{addr, \sigma_{Tx}})$ in the list $\mathcal{R}_1^\star$.
\end{itemize}

\smallskip
\noindent\hangindent 2em $\mathbf{\mathsf{O}^{owner\star}(\mathsf{quote\; generation;vk_{sign})}}.$ When $\mathcal{A}$ takes a query to the $\mathbf{\mathsf{O}^{owner}}$ oracle, $\mathcal{S}$ responds the same way as in the real world, except that $\mathcal{S}$ now stores all the $\mathsf{quote}$ corresponding to the user's queries in a list $\mathcal{R}_2^\star$. That is, $\mathcal{S}$ does the following algorithms.

\begin{itemize}
\item[-]  Call the $\mathbf{\mathsf{O}^{owner}}$ oracle with an input $(\mathsf{quote\; generation; vk_{sign}})$ and output $\mathsf{quote}$.
\item[-] Store $(\mathsf{vk_{sign}, quote})$ in the list $\mathcal{R}_2^\star$.
\end{itemize}

\smallskip
\noindent\hangindent 2em $\mathbf{\mathsf{O}^{delegatee\star}(\mathsf{key \;provision} ;\mathsf{quote})}.$ When $\mathcal{A}$ takes a query to the $\mathbf{\mathsf{O}^{delegatee}}$ oracle, $\mathcal{S}$ responds the same way as in the real world, except that $\mathcal{S}$ now stores all the $\mathsf{quote}$ corresponding to the user's queries in a list $\mathcal{C}^\star$. That is, $\mathcal{S}$ does the following algorithm.

\begin{itemize}
\item[-] Call $\mathbf{\mathsf{O}^{delegatee}}$ oracle with an input $(\mathsf{key \;provision} ;\mathsf{quote})$ and output $\mathsf{ct_{r}}$.
\item[-] Store $\mathsf{(quote,ct_{r})}$ in the list $\mathcal{C}^\star$.
\end{itemize}

%----------------------------------------------
\smallskip
For the PPT simulator $\mathcal{S}$, we prove the security by showing that the view of an adversary $\mathcal{A}$ in the real world is computationally indistinguishable from its view in the ideal world. Specifically, we establish a series of \textbf{Hybrids} that $\mathcal{A}$ cannot be distinguished with a non-negligible advantage as follows.

%----------------------------------------------
\medskip
\noindent\textbf{Hybrid 0.} $\mathsf{Exp^{real}_{DelegaCoin}(1^\lambda)}$ runs. 

%----------------------------------------------
\smallskip
\noindent\textbf{Hybrid 1.} As in \textit{Hybrid 0}, except that $\mathbf{KeyGen^{\star}(1^\lambda)}$ run by $\mathcal{S}$ is used to generate secret keys instead of
$\mathbf{KeyGen(1^\lambda)}$.

\begin{prf}
The proof is straightforward, storing corresponding answers in lists does not affect the view of $\mathcal{A}$. Thus,
$\textit{Hybrid 1}$ is indistinguishable from $\textit{Hybrid 0}$. 
\qed \end{prf}

\smallskip
\noindent\textbf{Hybrid 2.} As in \textit{Hybrid 1}, except that $\mathcal{S}$ maintains a list $\mathsf{\mathcal{C}^{\star}}$ of all  $\mathsf{quote =(hdl,tag_P,in,out,\sigma)}$ output by $\mathsf{HW.Run\&Quote(hdl_{\mathcal{O}},in)}$. And, when $\mathsf{HW.QuoteVerify(hdl_{\mathcal{D}}, pms,quote)}$ is called, $\mathcal{S}$ outputs $\bot$ if $\mathsf{quote \notin \mathcal{R}_2}$. ($\mathcal{R}_2$ is a quote returned by the real-world oracles that $\adv$ has queried as defined in Section~\ref{oracles}).

\begin{prf} If a fake quote is produced, then the step $\mathsf{HW.QuoteVerify(hdl_{\mathcal{O}}, pms,quote)}$ in the real word would make it output $\bot$. Thus, $\textit{Hybrid 2}$ differs from $\textit{Hybrid 1}$ only when $\mathcal{A}$ can produce a valid $\mathsf{quote}$ without knowing $\mathsf{sk_{\mathcal{O}}}$. Assume that there is an  adversary $\mathcal{A}$ can distinguish between $\textit{Hybrid 2}$ and $\textit{Hybrid 1}$. Obviously, this can be transformed to the ability against Remote Attestation as in Definition~\ref{remoteAttestation}. However, our assumption relies on the fact that the security of Remote Attestation holds. Thuerefore, \textit{Hybrid 2} is indistinguishable from \textit{Hybrid 1}.  \qed
\end{prf}
\smallskip
\noindent\textbf{Hybrid 3.} As in \textit{Hybrid 2}, except that when the $\mathbf{\mathsf{O}^{delegatee}}$ oracle calls $ \mathsf{HW.Run(hdl_{\mathcal{D}}, }$ $ \mathsf{ (\text{``provision''}, \mathsf{quote,\mathsf{pk_{\mathcal{O}}}, pms}))}$, $\mathcal{S}$ replaces $\mathsf{ct_r}$ as an encryption of zeros $\mathsf{\mathsf{PKE}.Enc(pk_{\mathcal{O}},1^{|r|})}$.

\begin{prf} The IND-CCA2 challenger provides the challenge public key $pk_{\mathcal{O}}$, and an adversary $\adv$ provides two messages $\mathsf{r}$ and $1^{|\mathsf{r}|}$, and further, the challenge returns an encryption of $\mathsf{r}$ or an encryption of $1^{|\mathsf{r}|}$, which is represented $\mathsf{ct_{\star}}$.
$\mathcal{S}$ sets $\mathsf{ct_{\star}}$ as the real output $\mathsf{ct_{r}}$. For $\mathsf{ct_r} \in \mathcal{C}$,
$\mathcal{S}$ can use $\mathsf{O}^{\mathsf{delegatee}}$ as it used in the real world. However, For $\mathsf{ct_r} \notin \mathcal{C}$, $\mathcal{S}$ neither has the oracles nor has the $\sk_{\mathcal{O}}$. But, the decryption oracle offered by the IND-CCA2 challenger can be used for any $\mathsf{ct_r} \notin \mathcal{C}$. Under this condition, if $\mathcal{A}$ can still distinguish \textit{Hybrid 3} and \textit{Hybrid 2}, we can forward the answer corresponding to $\mathcal{A}$'s answer to the IND-CCA2 challenger. If $\mathcal{A}$ can
distinguish between these two hybrids with a non-negligible probability, the IND-CCA2 security of $\mathsf{PKE}$ (see Definition~\ref{ccapke}) can
be broken with a non-negligible probability. \qed
\end{prf}

\smallskip
\noindent\textbf{Hybrid 4.} As in \textit{Hybrid 3}, except that $\mathcal{S}$ maintains a list $\mathcal{R}_1^\star$ of all transaction signature $\mathsf{\sigma_{Tx}}$ output by $\mathbf{\mathsf{O}^{owner}(\mathsf{signature\; creation; addr})}$ for $\mathsf{addr} \in \mathcal{R}_1$. When  $\mathsf{b} \gets \mathsf{\mathsf{S}.verify^{B}(pk_{Tx},\sigma_{Tx})}$ is called $\mathcal{S}$ outputs $\bot$ if $(\mathsf{addr}, \mathsf{\sigma_{Tx}})$, as components of a $\mathsf{Tx}$, do not belong to $\mathcal{R}_1$. Namely, $\mathsf{(\mathsf{addr}, \mathsf{\sigma_{Tx}}) \notin \mathcal{R}_1}$.

\begin{prf} If a transaction is given with an invalid signature, then the step $\mathsf{\mathsf{S}.Verify^{B}( pk_{Tx},\sigma_{Tx})}$ in the real word would make it output $\bot$. Thus, $\textit{Hybrid 4}$ differs from $\textit{Hybrid 3}$ only when $\mathcal{A}$ can produce a valid signature on $\mathsf{addr}$ which has never appeared before in the communication between $\mathcal{A}$ and the oracles. Let $\mathcal{A}$ be an adversary who can distinguish $\textit{Hybrid 4}$ and $\textit{Hybrid 3}$. We use it to break the EUF-CMA~\cite{goldwasser1988digital} security of signature scheme $\mathcal{S}$. We get a verification key $\mathsf{pk_{Tx}}$ and an access to $\mathsf{\mathsf{S}.Sign(sk_{Tx},\cdot)}$ oracle from the EUF-CMA challenger. Whenever $\mathcal{S}$ signs a message using $\mathsf{sk_{Tx}}$, it uses the $\mathsf{\mathsf{S}.Sign(sk_{Tx},\cdot)}$ oracle.  Also, our construction does not need a direct access to $\mathsf{sk_{Tx}}$ sign; it is used only to sign messages for the oracle provided by the challenger. Now, if $\mathcal{A}$ can distinguish two hybrids, the only reason is that $\mathcal{A}$ generates a valid signature $\mathsf{\sigma_{Tx}}$. Then, we can send such signature as forgery to the EUF-CMA~\cite{goldwasser1988digital} challenger. \qed
\end{prf}

%----------------------------------------------
\noindent\textbf{Hybrid 5.} As shown in \textit{Hybrid 4}, except that when the $\mathbf{\mathsf{O}^{owner}}$ oracle calls the function $\mathsf{HW.Run(hdl_{\mathcal{O}}, (\text{``start delegation''},addr))}$, $\mathcal{S}$ replaces $\mathsf{Enc}$ with $\mathsf{Enc^{\star}}$.

\begin{lemma}\label{lemma1}
If symmetric encryption scheme $\mathsf{SE}$ is IND-CPA secure, \textit{Hybrid 5} is indistinguishable from \textit{Hybrid 4}.
\end{lemma}

\begin{prf}
Whenever $\mathcal{A}$ provides a transaction $\mathsf{Tx}$ of its choice, $\mathcal{S}$ replies with zeros, e.g., $\mathsf{\mathsf{SE}.Enc(1^{|r|})}$, which is shown as follows.

\vspace{10ex}
\begin{center}
\begin{gameproof}[nr=3,name=\mathsf{Hybrid },arg=(1^n)]
\gameprocedure{%
 \pcln  \text{\dots}  \\
 \pcln  \mathsf{\mathsf{ct_{tx}} \gets \mathsf{SE}.\mathsf{Enc(r,Tx)}}  \\
 \pcln  \text{\dots}  
 }
\gameprocedure{%
   \text{\dots}  \\
   \gamechange{$\mathsf{\mathsf{ct_{tx}} \gets \mathsf{SE}.\mathsf{Enc(1^{|r|},Tx)}}$}  \\
   \text{\dots}  
}
%\addstartgamehop{hint=\footnotesize some ingoing hint}
\addgamehop{4}{5}{hint=\footnotesize replace the encryption with zeros, nodestyle=red}
% \addendgamehop{hint=\footnotesize some outgoing hint}
\end{gameproof}
\end{center}

Assume that there is an adversary $\mathcal{A}$ that is able
to distinguish the environments of \textit{Hybrid 5} and \textit{Hybrid 4}. Then, we build an adversary $\mathcal{A}^\star$ against IND-CPA secure of $\mathsf{SE}$.  Given a transaction $\mathsf{Tx}$ , if $\mathcal{A}$ distinguishes the encryption
of $\mathsf{r}$ from the encryption of $1^{\mathsf{|r|}}$, we forward the corresponding answer to the IND-CPA challenger.  \qed

\end{prf}

\noindent\textbf{Hybrid 6.} As in \textit{Hybrid 5}, except that when the $\mathcal{A}$ calls $\mathsf{HW.Run(hdl_{\mathcal{O}}, (\text{``state seal''},addr))}$, $\mathcal{S}$ replaces $\mathsf{Enc}$ with $\mathsf{Enc^{\star}}$.

\begin{prf}
The Indistinguishability between $\textit{Hybrid 6}$ and $\textit{Hybrid 5}$ can be directly
reduced to the IND-CPA property of $\mathsf{SE}$, which is similar to the lemma~\ref{lemma1}\qed
\end{prf}

%====================================================
\section{Implementation}
\label{sec-implementation}
%====================================================

We implement a prototype with three types of entities: the owner node, the delegatee node, and the blockchain system. The owner node and the delegatee node are separately running on two computers. The codes of these nodes are both developed in C++ using the $\text{Intel}^\circledR$ SGX SDK 1.6 under the operating system of Ubuntu 20.04.1 LTS. For the blockchain network, we adopt the Bitcoin testnet~\cite{bitcointest} as our prototype platform. Specifically, we employ SHA-256 as the hash algorithm, and ECDSA~\cite{johnson2001elliptic} with \textit{secp256k1}~\cite{sec20002} as the initial setting to sign transactions, which is the same with Bitcoin testnet's configuration.

\smallskip
\noindent\textbf{Functionalities.} We emphasize two main functionalities in our protocol, including \textit{isolated transaction generation} and \textit{remote attestation}. The delegation inside TEEs has full responsibility to govern the behaviours of participants. In particular, TEEs first calls the function $sgx\_create\_enclave$ and $enclave\_init\_ra$ to create and initialize an enclave \textit{E}$_{\mathcal{O}}$. Then, it derives the transaction key $sk_{Tx}$ under the user's invocation. 

%----------------------------------------------
\begin{algorithm} 
\label{algorithm1}
\caption{Remote Attestation}
  \BlankLine
   \KwIn{$\mathsf{request(quote, pms)}$}
    \KwOut{$\mathsf{b=0/1}$ }

 \BlankLine
 \textbf{parse} the received $\mathsf{quote}$ into $\mathsf{hdl,tag_P,in,out,\sigma}$ \\
 \textbf{verify} the validity of $\mathsf{vk_{sign}}$ \\
  \textbf{run} the algorithm  $\mathsf{HW.quoteVerify}$ with an input  $\mathsf{(pms,quote)}$\\
 \textbf{verify} the validity of $\mathsf{quote}$  \\
 \textbf{return} the results $\mathsf{b}$ if it passes ($\mathsf{1}$), or not ($\mathsf{0}$) \\
\end{algorithm}

Next, the system generates a bitcoin address  and a transaction with calling the function
$create\_address\_from\_string$ and $generate\_transaction$ respectively. \textit{E}$_{\mathcal{O}}$ keeps $sk_{Tx}$ in its global variable storage and signs the transaction with it while calling $generate\_transaction$. The transaction can only be generated inside the enclave without exposing to the public. Afterwards, \textit{E}$_{\mathcal{O}}$ creates a quote by calling the function $ra\_network\_send\_receive$, and proves to the delegatee that 
its enclave has been successfully initialized and is ready for the further delegation.

%====================================================
\section{Evaluation}
\label{sec-evaluation}
%====================================================

In this section, we evaluate the system with respect to \textit{performance} and \textit{disk space}. To have an accurate and fair test result, we repeat the measure for each operation 500 times and calculate the average.

%----------------------------------------------
\subsection{Performance}

The operations of public key generation and address create cost approximately the same time. This is due to the reason that they are both based on the same type of basic cryptographic primitives. The operations of transaction generation, state seal, and transaction decryption spend more time than the aforementioned operations because they combine more complex cryptographic functions. We also observe that the enclave initiation spends much more time than (transactions) key pair generations. Fortunately, the time used on enclave initiation can be omitted since the enclave each time launches only once (one-time operation). The state update spends the lowest time since most of the recorded messages are overlapped without the changes and only a small portion of data requires an update. The operations of coin deposit and transaction confirmation depend on the configuration of the Bitcoin testnet, varying from 10+ seconds to several minutes. Furthermore, we attach the time costs of the \textit{state seal} operation under increased transactions in Figure.\ref{fig-test} (right column). The time consumption grows slowly because a large portion of transactions are processed in batch. Remarkably, it costs less than 25 millisecond to finish all operations of coin delegation, which is significantly lower than the online transaction of Bitcoin testnet. This indicates that our solution is efficient in transaction processing and practical coin delegation.

\begin{table}[!hbtp]
 \caption{The average performance of various operations} 
 \label{tab-test}
  \centering
 \resizebox{0.75\linewidth}{!}{
 \begin{tabular}{llr}
    \toprule
    \textbf{Phase} & \textbf{Operation} & \textbf{Average Time / ms}  \\
    \midrule
    \multirow{2}{*}{\textit{System setup}}  & Enclave initiation   & $ 13.18940 $\\
    & Public key generation (Tx) & $ 0.34223 $ \\
    & Private key generation (Tx) &  $0.01119 $ \\
    \cmidrule{1-2}
    \multirow{2}{*}{\textit{Coin deposit}} & Address creation & $0.00690 $  \\
    & Coin deposit  & $ -$  \\
    \cmidrule{1-2}
    \multirow{4}{*}{\textit{Coin delegation}} & Transaction generation &  $ 0.78565 $ \\
     & Remote attestation & $19.50990 $  \\
     & State update & $ 0.00366 $  \\
     & State seal & $ 5.43957  $  \\
    \cmidrule{1-2}
    \multirow{2}{*}{\textit{Coin spend}} & Transaction decryption & $ - $  \\
     & Transaction confirmation &  $ - $   \\
    \bottomrule
  \end{tabular}
  }
\end{table}

%----------------------------------------------
\subsection{Disk Space}
%---------------------------------------------- 

\begin{figure}[!hbt]
\centering
\caption{Used disk space and time consuming of state seal}
\includegraphics[width=0.65\textwidth]{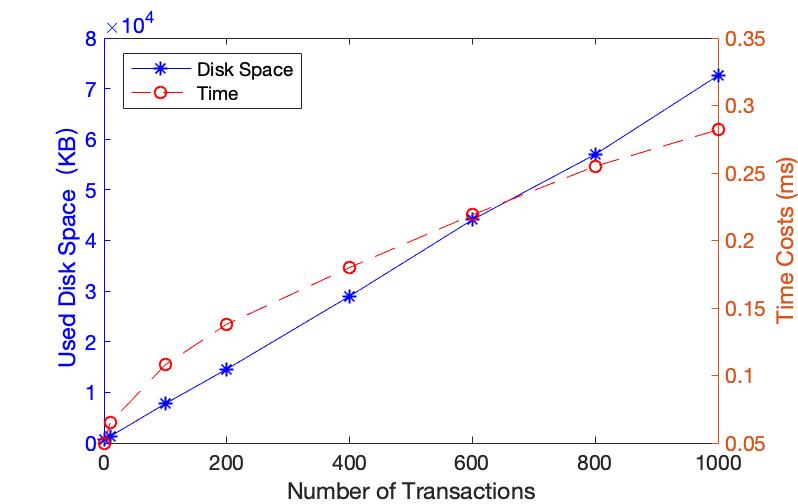}
\label{fig-test}
\end{figure}

In this part, we provide an evaluation of the disk space of the sealed state. We simulate the situation in DelegaCoin when more delegation transactions join the network. The transaction creation rate is set to be 560 transactions/second. We monitor space usage and the corresponding growth rate. Each transaction occupies approximately 700 KB of storage space. We test eight sets of experiments with an increased number of transactions in the sequence $1, 10, 100, 200, 400, 600, 800, 1000$.  The results, as shown in Figure.\ref{fig-test} (left column), indicate that the size of the disk usage grows linearly with increased delegation transactions. The reason is straightforward: the disk usage closely relates to the involved transactions that are stored in the list. In our configurations, the transaction generation rate stays fixed. Therefore, the used space is proportional to the increased transactions.

%====================================================
\section{Conclusion}
\label{sec-conclusion}
%====================================================

Decentralized cryptocurrencies such as Bitcoin~\cite{nakamoto2008bitcoin} provide an alternative approach for peer-to-peer payments. However, such payments are time-consuming. In this paper, we provide a secure and practical  TEEs-based offline delegatable cryptocurrency system. TEEs are used as the primitives to establish a secure delegation channel and offer better storage protection of metadata (keys, policy). An owner can delegate the coin through an offline-transaction asynchronously with the blockchain network. A formal analysis, prototype implementation and further evaluation demonstrate that our scheme is provably secure and practically feasible.

\textit{Future Work.} There is an insurmountable gap between the theoretical case and real application.  Although our scheme is proved to be theoretically secure, a lot of risks still exist in practical scenarios. The countermeasures to reduce these risks will be explored.

\smallskip
\noindent\textbf{Acknowledgments.} Rujia Li and Qi Wang were supported by Guangdong Provincial Key Laboratory (Grant No. 2020B121201001).

%-------------------------------------------------------------------------------
%-------------------------------------------------------------------------------
%-------------------------------------------------------------------------------
\normalem
\bibliographystyle{unsrt}
\bibliography{bib}

\newpage
%------------------------------------------------------
\appendix
\section*{Appendix A. Protocol Workflow}
\label{appendix:a}
%------------------------------------------------------
This part provides an overview of the transaction workflow. It assists to demonstrate the operating mechanism of a TEEs-enabled delegation scheme and its corresponding connections between the blockchain network and users.

\setlength{\belowcaptionskip}{-15pt}
\begin{figure}[hbt!]
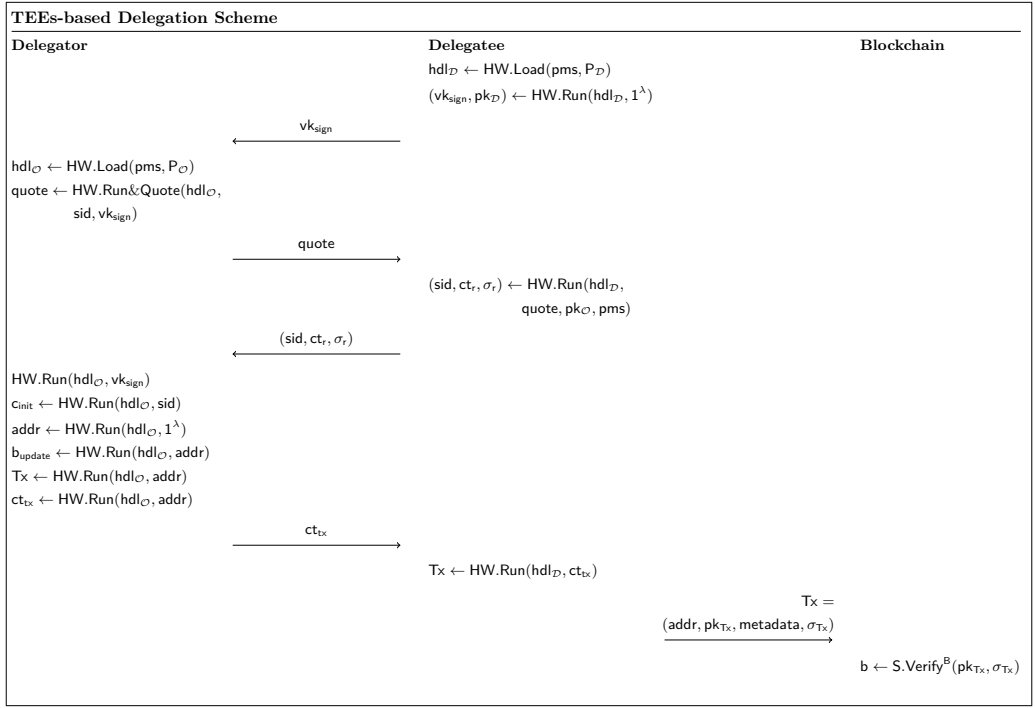

\resizebox{\linewidth}{!}{ 
\fbox{%
        \procedure{\textbf{TEEs-based Delegation Scheme}}{%
         \textbf{Delegator} \< \< \textbf{Delegatee} \< \< \textbf{Blockchain}  \\
         %------------------------------------------
         \< \< \mathsf{hdl_{\mathcal{D}}} \gets \mathsf{HW.Load(pms,P_{\mathcal{D}})} \\
         \< \< \mathsf{(\mathsf{vk_{sign}}, pk_{\mathcal{D}}) \gets HW.Run(hdl_{\mathcal{D}}, 1^\lambda)}\\
         %%%%%%%%%%%%%%%%%%%
         \< \sendmessageleft*{ \mathsf{vk_{sign}} } \<  \\
         \mathsf{hdl}_{\mathcal{O}} \gets \mathsf{HW.Load(pms,P_{\mathcal{O}})} \<\< \\
         \mathsf{quote \gets HW.Run\&Quote(hdl_{\mathcal{O}}, } \< \<  \\
         \quad\quad\quad\quad \mathsf{sid, vk_{sign}}) \< \<  \\
         %%%%%%%%%%%%%%%%%%%
         \< \sendmessageright*{\mathsf{quote}} \<   \\
         \< \< \mathsf{(sid, ct_{r}, \sigma_{r})} \gets \mathsf{HW.Run(hdl_{\mathcal{D}}, } \\
         \< \< \quad\quad\quad\quad\quad\quad \mathsf{quote,pk_{\mathcal{O}}, pms}) \\
         %%%%%%%%%%%%%%%%%%%
         \< \sendmessageleft*{\mathsf{(sid, ct_{r}, \sigma_{r})}} \<  \\
         \mathsf{HW.Run(hdl_{\mathcal{O}}, \mathsf{vk_{sign}})} \< \<  \\
         \mathsf{c_{init} \gets HW.Run(hdl_{\mathcal{O}}, sid)}  \< \<  \\
         \mathsf{addr \gets HW.Run(hdl_{\mathcal{O}}, 1^\lambda)}  \< \<  \\
         \mathsf{b_{update} \gets HW.Run(hdl_{\mathcal{O}}, addr)} \< \<  \\
         \mathsf{Tx \gets HW.Run(hdl_{\mathcal{O}}, addr)} \< \<  \\
         \mathsf{ct_{tx} \gets HW.Run(hdl_{\mathcal{O}}, addr)} \< \<  \\
         %%%%%%%%%%%%%%%%%%%
         \<  \sendmessageright*{\mathsf{ct_{tx}}}  \<    \\
         \< \<  \mathsf{Tx \gets HW.Run(hdl_{\mathcal{D}}, \mathsf{ct_{tx}})} \\
         %%%%%%%%%%%%%%%%%%%
         \<\<\<  \sendmessageright*{\mathsf{Tx = }\\ \mathsf{(addr,pk_{Tx},metadata,\sigma_{Tx})}}  \<    \\
        \<\<\<\< \mathsf{b} \gets \mathsf{\mathsf{S}.Verify^{B}(pk_{Tx},\sigma_{Tx})}\\
        }
        }      
        } %resize box to fit linewidth 
        \caption{The Transaction Flow of Delegation}
\end{figure}

\begin{comment}
\begin{figure}[htb!]
    \centering
     \fbox{
        \procedure{}{\\
         \begin{subprocedure}\procedure{\textbf{Issuer(tracer)}}{y_t,\zeta_1}\end{subprocedure} \< \< 
         \begin{subprocedure}\procedure{\textbf{Blockchain}}{y_t}\end{subprocedure}  \pclb
        \pcintertext[dotted]{build a secure channel} \\
        \< \sendmessageright*{\zeta_1} \<  \\
         \<\< \dbox{\begin{subprocedure}\procedure{}{
         \< \< I_{id} = \zeta_1^{1/x_t} = z_1^{\lambda/x_t} \\
         \< \< = y_t^{\upsilon\lambda/x_t} = g ^{\upsilon\lambda} = \xi^{\upsilon} 
        }\end{subprocedure}}  \\
        \< \sendmessageleft*{\xi^{\upsilon}} \<   \\
                \begin{subprocedure}\procedure{}{
        (\zeta_1,\xi^{\upsilon})   \< \<
        }\end{subprocedure}  \< \< \\
         \<\< }
         }
        \caption{The identity tracing protocol.}
        \label{fig:identity}
\end{figure}

\end{comment}

%------------------------------------------------------
\section*{Appendix B. Resource Availability}
\label{appendix:b}
%------------------------------------------------------

\begin{itemize}
\item Implementation Code: \url{https://github.com/TEEs-projects/DelegaCoin}

\item Test Data: \url{https://github.com/TEEs-projects/DelegaCoin/tree/main/test_data}

\item NDSS Poster Version: \url{https://www.ndss-symposium.org/wp-content/uploads/NDSS2021posters_paper_14.pdf}

\item ICBC Conference Version: \url{https://icbc2021.ieee-icbc.org/program}

\end{itemize}

\newpage
%------------------------------------------------------
\section*{Appendix C. Notations}
\label{appendix:c}
%------------------------------------------------------
\begin{table*}[!hbt]
 \caption{Featured Notations}\label{tab-notation}
 \label{node}
  \centering
    \resizebox{\linewidth}{!}{  
    \begin{tabular}[t]{cll}
    \toprule
    \textbf{ Symbol }  & \quad \textbf{Item}  & \quad\quad\quad\quad\quad\quad \textbf{ Functionalities} \\  \midrule
     $\mathcal{O}$ &  Delegator & also known as coin owner, the person who sends the coins  \\  \midrule
     $\mathcal{D}$  & Delegatee \, & the person who receives the coins\\  \midrule
     $\mathcal{B}$  & Blockchain & an ideal blockchain environment provides all types of basic functions \\ \midrule
     $\mathsf{Tx}$  & Transaction \, & the transaction in blockchain network \\  \midrule
     $E_{\mathcal{O}}/E_{\mathcal{D}}$  & Enclave \, & the Delegatee's/Delegator's Intel enclave instance \\ \midrule
     $\mathsf{TEE}$  & TEEs   & a real TEEs environment, sometimes used in superscript for indication \\ \midrule
     $\mathsf{hdl}$ & Handle  & an intermediate parameter when initiating  TEEs \\ \midrule
     $\mathsf{quote}$ & Quote &  a flag to request operations when running TEEs \\ \midrule
     $\mathsf{pms}$ & Parameters  &  intermediate parameters when running  TEEs \\ \midrule
     $\mathsf{pk/sk}$ & Key pair & the public key and private key to encrypt/decrypt states \\ \midrule
     $\mathsf{vk/sk_{sign}}$ & Key pair & the key pair to identify a specific entity (delegatee) \\ \midrule
     $\mathsf{key_{seal}}$ & Private key  & a sealing key used to export the state to the trusted storage  \\ \midrule
     $\mathsf{r}$ & Private key  & a symmetric encryption key $r$   \\ \midrule
     $\mathsf{ct_{r}}$ & Ciphertext & the ciphertext under a symmetric encryption key with $r$ inside TEEs \\ \midrule
     $\mathsf{b}$ & Account balance & the subscript $\mathsf{init/deposit/update}$ means the status in different stages  \\ \midrule
     $\mathsf{c}$ &   Encrypted balance &  the balance that has been encrypted and transferred \\ \midrule
     $\sigma$ & Signature & a valid signature, the subscript indicates its corresponding signer  \\ \midrule
     $\mathsf{HW}$ & Hardware  & a ideal and secure hardware functionality used in proofs \\ \midrule
     $\mathcal{P}$ & Program space  & a program that contains a set of algorithms, instantiated as $P$ \\\midrule
     $\mathsf{O}$  & Oracle  & an environment that can provide ideal functionalities \\ \midrule
     $\mathcal{U}(\cdot)$ &  Oracle & a universal oracle can provide simulated answers \\ \midrule
     $\mathcal{S}$ & Simulator  & an ideal environment that can simulate some behaviours  \\ \midrule
     $\mathcal{A}$ & Adversary &  an adversary who has some ability to launch attacks \\ \midrule
     $\mathsf{\lambda}$  & Security parameter & a type of parameter to adjust the security level of algorithms \\ \midrule
     $\mathsf{negl(\lambda)}$ &  Negligible function & a function to show the negligible differences in security proofs \\ \midrule
     $\mathsf{Exp}$ & Experiment &  an experiment that show the game and operations in proofs \\ \midrule
     $\mathsf{PKE}$ & Algorithm &  an IND-CCA2 secure public key encryption scheme \\ \midrule
     $\mathsf{S}$ & Algorithm &  an existentially unforgeable (EUF-CMA) signature scheme \\ \midrule
     $\mathsf{SE}$ & Algorithm & a IND-CPA secure symmetric encryption scheme \\
    \bottomrule
    \end{tabular}
    }
\end{table*}

\end{document}